%% file: main.tex
\documentclass[acmsmall]{acmart}

\usepackage{csquotes} 
\usepackage{longtable}
\usepackage{soul} 
\usepackage{subcaption}
\usepackage{graphicx}
\usepackage{makecell}

\AtBeginDocument{%
  }

\setcopyright{cc}
\setcctype{by}
\acmJournal{PACMHCI}
\acmYear{2025} \acmVolume{9} \acmNumber{2} \acmArticle{CSCW067} \acmMonth{4} \acmPrice{}\acmDOI{10.1145/3710965}

\begin{document}

\title[{\textit{Engage} and \textit{Mobilize} for Emergency Management}]{
\textit{Engage} and \textit{Mobilize}! Understanding Evolving Patterns 
of Social Media Usage in Emergency Management}


\author{Hemant Purohit}
\email{hpurohit@gmu.edu}
\affiliation{%
  \institution{George Mason University}
  \city{Fairfax}
  \country{USA}
} 

\author{Cody Buntain}
\email{cbuntain@umd.edu}
\affiliation{%
  \institution{University of Maryland}
  \city{College Park}
  \country{USA}
}

\author{Amanda Lee Hughes}
\email{amanda_hughes@byu.edu}
\affiliation{%
  \institution{Brigham Young University}
  \city{Provo}
  \country{USA}
}

\author{Steve Peterson}
\email{stevepeterson2@gmail.com}
\affiliation{%
  \institution{Montgomery County Community Emergency Response Team}
  \city{Gaithersburg}
  \country{USA}
}

\author{Valerio Lorini}
\email{valerio.lorini@europarl.europa.eu}
\affiliation{%
  \institution{European Parliament}
  \city{Luxembourg}
  \country{Luxembourg}
}

\author{Carlos Castillo}
\email{carlos.castillo@upf.edu}
\affiliation{%
  \institution{Universitat Pompeu Fabra}
  \city{Barcelona}
  \country{Spain}
}

\renewcommand{\shortauthors}{Purohit et al.}

\newcommand{\inblue}[1]{{\color{black}{#1}}}
\newcommand{\inbrown}[1]{{\color{black}{#1}}}

\begin{abstract}
The work of Emergency Management (EM) agencies requires timely collection of relevant data to inform decision-making for operations and public communication before, during, and after a disaster. However, the limited human resources available to deploy for field data collection is a persistent problem for EM agencies.  
Thus, over the last decade, many of these agencies have started leveraging social media as a supplemental data source and a new venue to engage with the public. 
Such uses present both opportunities and challenges. 
While prior research has analyzed the potential benefits and attitudes of practitioners and the public when leveraging social media during disasters, a gap exists in the critical analysis of the actual practices and uses of social media among EM agencies, across both geographical regions and phases of the 
EM lifecycle - typically mitigation, preparedness, response, and recovery.     
In this paper, we conduct a mixed-method analysis to update and fill this gap on how EM practitioners in the U.S. and Europe use social media, 
building on a survey study of about 
150 professionals and a follow-up interview study with 11 participants. 
The results indicate that using social media is no longer a non-traditional practice in operational and informational processes for the decision-making of EM agencies 
\inblue{working at both the local level (e.g., county or town) and non-local level (e.g., state/province, federal/national) for emergency management. 
Especially, the  practitioners affiliated with agencies working at the local level have a very high perceived value of social media for situational awareness (e.g., analyzing disaster extent and impact) and public communication (e.g., disseminating timely information and correcting errors in crisis coverage).}    
Further, practitioners now engage with the public during the preparedness phase 
to mobilize them during the response phase. 
We present a model to understand the current practices of communication between  agencies and the public, as well as among practitioners while leveraging social media. 
\inblue{We also discuss novel challenges,} including public fragmentation caused by the increasing use of multiple social media platforms, information integrity, and social listening expectations. 
We conclude with the policy, technological, and socio-technical needs to design future social media analytics systems to support the work of EM agencies in such communication.     
\end{abstract}




\maketitle

\input{01-intro}

\input{02-relwork}

\input{03-methods}


\input{04-results-survey}


\input{05-results-qual-analysis}



\input{06-discussion}

\section{Conclusions} \label{sec:conclusions}

This paper studies the evolution of the use of social media by practitioners 
for a variety of work processes at EM agencies. 
We conducted a mixed-method analysis via a survey study and a follow-up interview study with practitioners across the U.S. and Europe with reference to the growth period of social media platforms over the last two decades. 
We found that using social media is now considered part of the traditional practice in various work functions at EM agencies \inblue{(both at the local and non-local levels)}, such as public communication. 
Practitioners have adopted social media in their roles and responsibilities across different phases of the EM lifecycle, contrary to the extensive focus on the response phase in the research literature.   
We observed that the cooperation practices between practitioners at official EM agencies and community organizations have evolved to take advantage of social media-based credibility for building trustworthy relationships and leveraging it to address the well-known barrier of limited resources in collecting and processing social media data.  
We further discovered novel challenges for EM agencies in using social media, including fragmentation across platforms that require 
inclusive, cross-platform technology solutions; information integrity requirements; lack of moderation capabilities; engagement metrics for reach and inclusivity; social listening expectations of the public; multimodal communication expectations, etc.   
Lastly, we have presented novel insights on policy, technological, and socio-technical design needs to support the work of EM agencies in the future, such as cross-platform, cross-language, and cross-modal analytics tools to support public communication, diversity in community engagement, procedures and policies for emergency communication and mobilization, and so on. 

\subsection{Limitations and Future Work}

\inblue{This work is novel in bridging experiences and insights among EM practitioners across multiple regions, countries, and organizational structures.} 
\inblue{At the same time, this breadth introduces notable limitations that should be addressed in future studies of EM practitioners' use of new media and emerging technologies such as AI tools, which we outline below:}

\paragraph{Cross-Cultural and Cross-National Processes} Designing and conducting a mixed-method analysis with participants working across geographies with different cultural and communication norms is a difficult task. \inblue{This work attempts to} 
harmonize the terminology and concepts for work processes (e.g., geographical units as county versus district or state versus province) when designing the survey and interview questionnaires. 
\inblue{While this harmonization was done in consultation with} and involving practitioners and researchers across the U.S. and Europe, \inblue{as we demonstrate above in \S5, a variety of language has been used to define similar concepts in the EM context, and} we might not have covered the conceptual terms as expected by practitioners in a specific region. 
We also acknowledge the existence of participation bias  in the surveys \inblue{and sample size limitations in interviews}, as typically observed in  \inblue{similar studies} in prior research. 
We tried to limit these biases by disseminating surveys across as many diverse professional groups in the EM domain as we could \inblue{and requesting participation in the follow-on interviews by highlighting the impact of such studies 
in advancing emergency management}. 
Resources and practices developed in this study could provide a foundation for future work along this direction, \inblue{potentially engaging with international standards bodies and EM associations like the IAEM}.

\paragraph{Social Media in Broader EM Contexts} \inblue{We note that our} survey and interviews were designed around the two key work functions of EM agencies where we have observed extensive literature in crisis informatics: 1) collection practices for situational awareness to aid operational decision-making, and 2) dissemination practices for public communication. 
\inblue{As documented in FEMA's emergency support functions (ESFs), these functions alone do not cover the full breadth of EM practices, as we in fact discuss above in our findings about the increasing use of social media for preparedness and audience-building functions.} 
\inblue{As such, we encourage} future studies \inblue{to} expand investigations into social media across different phases of the EM lifecycle, especially in the mitigation and preparedness.

\paragraph{EM in a Post-LLM World} \inblue{Owing to the time-intensive process of interviewing and qualitative study,} when we designed \inblue{and fielded our survey and interviews}, OpenAI's ChatGPT \cite{chatgpt}  
had not yet introduced general audiences to \inblue{widely available and usable} generative AI capabilities. 
\inblue{Consequently,} participant responses to questions like Q10.2 on tools for public communication, may now \inblue{be quite different owing to the much wider availability of LLM-based and generative-AI tools}. Future studies \inblue{should investigate this potential} value of emerging AI tools in aiding EM agencies across all phases of the EM lifecycle, \inblue{as many of the technology design needs we identify in \S6.3 may be much more easily facilitated given these new tools.}

\paragraph{An International Perspective, but not a Global One} \inblue{We acknowledge the limitations that come with} our study's focus on the U.S. and Europe.
\inblue{We therefore also acknowledge that} our insights on the current practices of EM practitioners \inblue{likely do} not apply worldwide. 
\inblue{This limited representation is especially true with respect to} the adoption of social media, as \inblue{social media adoption in the U.S. and Europe has outpaced} other parts of the world. 
\inblue{Despite this limitation,} social media usage by practitioners at EM agencies needs to be better understood, \inblue{as we touch upon in our discussion of coordination between Portuguese organizations and civilians in Mozambique.}  
\inblue{We therefore encourage future cross-national research, especially related to low- and middle-income countries, where repurposing new media infrastructure and engagement with European and U.S.-practitioners might be particularly impactful}. 

\inblue{
This need for cross-national research is especially salient when we discuss the use of AI in EM practices. 
AI use varies greatly across countries, as evidenced by the variety of governmental approaches to AI use--e.g., the U.S. executive order on and inventory of AI in government \cite{gov_ai_use_US};  
the UK National Audit Office's report of AI in UK government functions \cite{gov_ai_use_UK};  
and the EU's own overview of AI in public services~\cite{misuraca2020ai}. 
How these various regulations, reports, and inventories will be integrated into public organizations' EM practices remains to be seen, and we may instead see private and civilian-led organizations make more use of the AI tools while public organizations wrestle with these policy questions.\\ 
}

\noindent \textit{\textbf{Acknowledgment.}} Authors thank emergency management professionals (anonymized for ethics protocol) for their invaluable support in this study, as well as Jorge Gomes, Alexandre Penha, Paola Rufolo, and the Copernicus Emergency Services team at the Joint Research Centre of the European Commission for insightful discussions during the planning of this research study.  

\bibliographystyle{ACM-Reference-Format}
\bibliography{bibliography.hughes,bibliography.buntain,bibliography.purohit,bibliography.chato,bibliography.lorini}

\newpage
\appendix
\begin{center}
    \textbf{\inblue{APPENDIX}}
\end{center}

\begin{longtable}{
lp{0.9\textwidth}
} 
\caption{\inblue{Survey questions.}}\label{table:survey-questions}\\ 
 \hline
 & \textbf{Question}  
  \\ [0.5ex] 
 \hline\hline
 1.& 
 \textit{Where do you work?} 
   (drop-down list for Countries)
  \\
  \\
 2.& 
 \textit{What discipline does your agency represent?} 
 (multiple answers accepted): Fire service, Emergency medical service/Ambulance service, 
 Telecommunications, Public works (water, sanitation), Transportation, Health/hospitals, Police/security, 
 Emergency management, Non-governmental organizations (NGOs), Other (Comment Text-box)
 \\
 \\
 3.& 
 \textit{What is your position?} 
 (Comment Text-box) - e.g., Firefighter, Fire Chief, Emergency Medical Technician, Public 
 Information Officer,  Police Chief, Director, etc. 
 \\ 
 \\
 4.& 
 \textit{Please describe your role/responsibility in one or two lines.} 
 (Comment Text-box) 
 \\ 
 \\
 5.& 
 \textit{At what level does your agency work for emergency management?} 
 (multiple answers accepted): 
 \{Choices for all countries excluding North America continent: International, National/ 
 Federal, Provincial/ Regional/District, Municipal/Town/Local, Other (Comment Text-box)\}, 
 \{Choices for North America continent-based countries: International, Federal, State, 
 County/Local, Other (Comment Text-box)\} 
 \\
 \\
 6.& 
 \textit{How often does your agency use social media?} 
 (multiple choice question): 
 (Daily, Weekly, Reactively (incident-based), Never) 
 \\
 \\
 7.& 
 \textit{How often do you use social media in a personal capacity?} 
 (multiple choice question): 
 (Daily, Weekly, Reactively (incident-based), Never) 
 \\
 \\
 8.& \textit{What situational awareness value do you think social media could provide to your agency?} 
 (Likert Scale Rating) -- Disaster extent and magnitude [SCALE 1-5], Analyzing disaster impact for infrastructure/population/services [SCALE 1-5], Analyzing resource needs of victims [SCALE 1-5], Authoritative information about response of other official agencies [SCALE 1-5]
 \\
 \\
 9.& \textit{What public communication value do you think social media could provide to your agency?}  
 (Likert Scale Rating) -- Engage and coordinate with media organizations for public outreach [SCALE 1-5], Coordinate and disseminate accessible, timely information [SCALE 1-5], Correct potential errors in crisis coverage [SCALE 1-5], Engage with members of the public - respond to questions and concerns [SCALE 1-5] 
 \\
 \\ 
 \hline 
& \textit{The following questions are only asked if the answer to Q6 is not `never':}  
 \\ 
 \hline 
 10.1. & \textit{What type of tools does your agency currently use to gather incident-specific information?} 
 -- Please list: [Text box] 
 \\
 \\
 10.2.& \textit{What type of tools does your agency currently use to publicize incident-specific information?}  
 -- Please list: [Text box] 
 \\ 
 \\
 10.3.& \textit{Using the above-listed tool(s), what does your agency use social media for?} 
 (multiple answers accepted), Continuous monitoring, Incident-based monitoring, Complementing previously known information, Public communication / community management including correcting rumors, Other (please explain) - [Text box]
 \\
 \\
 10.4.& \textit{What do you consider the most critical barrier(s) to emergency management agencies not using social media in an emergency operations center environment?} 
 (multiple answers accepted) -- Unverified information or sources, Lack of human resources and management support, Complexity of automated tools [comment on tools: Text-box], Lack of training for using automated tools [comment on tools: Text box], Poor quality of information processing by automated tools [comment on tools: Text box], Other (please explain) - [Text box] 
 \\ 
 \\
 11.& \textit{Would you like to volunteer to further contribute to the study by participating in a focus group in the coming months?} 
 (Note: Future focus group participants will be recognized with Certificates of Appreciation by the Copernicus Emergency Services from the European Commission 
 Joint Research Centre
 ) --Yes [please provide email: [Text box]], No [Would you be interested in seeing the results of the analysis? If so, kindly provide email to send the report to: [Text box]]
 \\ [1ex] 
 \hline
\end{longtable}

\newpage 

\begin{longtable}{
lp{0.9\textwidth}
} 
\caption{\inblue{Interview questions.}}\label{table:interview-questions}\\ 
 \hline
 & \textbf{Scenario}  
  \\ [0.5ex] 
 \hline 
 & 
 Recall a disaster event that you (or your agency) experienced where you obtained or searched for information 
 pertaining to the event (e.g., number of victims, road closures, damage impact). 
  \\ [0.5ex] 
  \hline
  & \textbf{Questions}  
  \\ [0.5ex] 
 \hline 
 1.& 
 What type of event have you been involved in? What happened? What was your role (e.g., unit lead, 
 administrative, decision-maker) and the level (e.g., local, regional, state) of agency? Was there 
 mutual aid or collaboration procedure? Was an Emergency Operations Center (EOC) activated? 
 \\
 \\
 2a.& 
 What kind of situational awareness information did you and/or your agency search for and how? 
 If you did not attain any from social media, what information would you deem relevant to 
 situational awareness? 
 \\ 
 \\
 2b.& 
 Please describe relevant disaster event information you or your agency found. Was it found through 
 monitoring social media? If not, do you think that type of information could possibly be found on social 
 media? How? 
 \\
 \\
 2c.& 
 What challenges did you face when searching for information through social media? 
 \\
 \\ 
 2d.& 
 What kinds of information did you or your agency communicate to the public and how? 
 Which traditional and non-traditional tools did you use? 
 \\
 \\ 
 2e.& 
 If you used social media, what types of information did you share? What platforms did you use? 
 Did you engage with the public? 
 \\
 \\ 
 2f.& 
 What worked well when using social media to disseminate information? What did not (e.g., any challenges)? 
 \\
 \\
 3.& 
 Describe an instance where social media provided a unique insight that you would have likely missed 
 without access to social media? 
 \\
 \\ 
 4.& 
 Does your agency communicate using social media with its neighboring agencies during disaster events? 
 \\
 \\ 
 5.& 
 Do social media feeds reside within your EOC? Is social media integrated within your communication plan 
 or ConOps? Does your agency have a Standard Operating Procedure (SOP) for social media use during 
 disasters? 
 \\
 \\ 
 6.& 
 What other thoughts would you like to share? 
 \\ [1ex] 
 \hline
\end{longtable} 

\end{document}

%% file: 01-intro.tex
\section{Introduction}

Much has been written about how to use social media platforms in emergency management (EM) before, during, and after disaster events. 
Early research focused on using Twitter \inbrown{(now X.com)} to improve situational awareness \cite{10.1109/mis.2012.6.2012,vieweg_microblogging_2010}, but over the years, the field of ``crisis informatics''~\cite{10.1126/science.aag2579.2016ogo} has branched in many directions, as attested by various surveys spanning multiple editions and years \cite{10.1145/2771588.2015,palen_social_2018, 10.1016/j.ijinfomgt.2019.04.004.2019,purohit_peterson2020social,10.1016/j.ijdrr.2020.101584.2020jd} (see Section~\ref{sec:relwork} for additional references).
Despite this abundance of research, EM practitioners have not fully realized 
the expected transformations from leveraging social media \cite{DHS_ST2014} in preparedness, response, and recovery phases of EM for which researchers have strived \cite{reuter.iscram:2018}. 
Researchers and practitioners have cited numerous reasons for this limited impact, including an over-reliance on Twitter/X data analysis \cite{10.1126/science.aag2579.2016ogo,reuter.iscram:2018}, concerns of information quality \cite{lorini2021social, hiltz_exploring_2020}, information overload \cite{lorini2021social}, organizational policies that discourage or prevent social media use \cite{hughes_evolving_2012,hiltz_exploring_2020}, and limited tools to handle visual media \cite{lorini2021social}, among others. 
However, most of the proposed solutions to these issues have focused on technical offerings that alleviate data-related problems. 
This ``technosolutionism'' or ``technochauvinism'' \cite{broussard2018artificial} has motivated 
critical questions, including how EM 
 practitioners have adopted social media for their needs over time, and how they have themselves adapted to the public's use of these platforms. 

To address these questions, this paper takes a 
comprehensive 
approach to understand the current practices of 
EM practitioners to incorporate social media into their socio-technical processes, and how these processes have evolved. 
Specifically, we employ a mixed-method approach in this study. 
We design and conduct 
a survey for practitioners
across various countries, and engage a subset of them in interviews to learn how they 
see and use social media in their work. 
We find new processes wherein practitioners have integrated social media into their outreach plans and performance indicators for public engagement, as well as how they actively use such virtual platforms to mobilize and collaborate with the public, locally, nationally, and internationally. 
We present a model of the four primary communication channels among agencies and the public. This model is then used to organize our findings and 
understand the differences in the current practices of agencies across key operational and informational processes such as data collection and public engagement. 
Our analysis 
validates long-standing obstacles for EM agencies in leveraging social media. For instance, dealing with incomplete information, 
determining exact locations of events referenced within social media messages (geolocating), and understanding the motivations and intentions of message posters and audiences. 
We further discover 
new issues around policies governing current practices of EM, platform-usage fragmentation among the public, 
and other barriers that technical solutions alone may be unable to address.

Building on these confirmatory and revelatory findings, this paper contributes an updated and more complete understanding of EM practitioners' current use of social media 
across different EM phases and also, discusses policy and socio-technical needs for the future. 
Section~\ref{sec:relwork} describes previous work related to ours and Section~\ref{sec:methods} presents our methodology, which includes a survey and an interview study. 
Sections~\ref{sec:findings-survey} and \ref{sec:findings-focus}, respectively, present the insights obtained from the survey and interviews with practitioners.
Section~\ref{sec:discussion} presents a discussion of these findings and Section~\ref{sec:conclusions} our conclusions. 

%% file: 02-relwork.tex
\section{Related Work}\label{sec:relwork}

This work builds on and makes contributions to a community that spans a broad spectrum of researchers, from social science and computational social science researchers studying how social media is used during crises, to computer scientists and developers building new crisis informatics tools to facilitate social media's use among EM practitioners.

As early as 2007, \citet{10.1145/1240624.1240736.2007} outlined how information \& communication technology ``will give further rise to \emph{improvised} activities'' during crises.
They also rightly foresaw how formal response organizations should respond to these activities, even before social media platforms were largely discussed--in fact, their paper does not mention explicitly ``social media'' at all.
Over time, such research has coalesced into the interdisciplinary field of ``crisis informatics,'' at the intersection of human-computer interaction and social science research \cite{10.1126/science.aag2579.2016ogo}.
Crisis informatics research continues to evolve as well, as social media platforms and EM practitioners' uses of these spaces expand 
and change over time.

Consequently, surveys of the field now span multiple editions and years--e.g., Imran et al. in 2015 \cite{10.1145/2771588.2015}, Castillo in 2016 \cite{castillo2016big}, Palen and Hughes in 2018 \cite{palen_social_2018}, Reuter and Kaufhold in 2018 \cite{reuter2018fifteen}, Zhang et al. in 2019 \cite{10.1016/j.ijinfomgt.2019.04.004.2019}, and Saroj and Pal in 2020 \cite{10.1016/j.ijdrr.2020.101584.2020jd}, etc. 
\inblue{Yet even since the most recent survey paper cited here from 2020, significant changes have occurred. For instance, Twitter rebranded to X and restricted free access to their API for data collection, complicating research  and practitioner efforts. Advances in machine learning and generative AI have provided new opportunities for emergency managers to make sense of large-scale social media datasets.}
It is to this evolving understanding that this work contributes, as these online social spaces, their audiences, their owners, technologies, and the policy environment, all \inblue{continue to see rapid growth and change}.

\subsection{Tools for Processing Social Media Data}

Much of the technical work in this space has focused on extracting information from social media. 
To a large extent, it is  done with the purpose 
of mitigating the information overload for practitioners that results from directly consuming social media postings related to a developing event. 

In the surveys cited previously, we find hundreds of papers describing various mechanisms by which human operators are presented with messages that have been automatically filtered, prioritized, organized for aligning emergency services and volunteer activities, or summarized. 
Examples include, among many others, methods to support crisis volunteers~\cite{10.1145/2702123.2702171.2015,cobb_designing_2014} and their task coordination to aid EM agencies~\cite{ludwig2015crowdmonitor,purohit2014identifying}, or to surface actionable and serviceable requests \cite{zade_situational_2018,10.1109/asonam.2018.8508709.2018,li2021hello}, or to summarize~\cite{vitiugin2022cross,mccreadie2023crisisfacts}. 
In general, this class of systems can be broadly described as human-centered, human-in-the-loop solutions~\cite{monarch_human_in_the_loop_2021} for information filtering, and they are powered by natural language processing (NLP), computer vision (CV), network-based methods, or more recently, methods that may combine multiple modalities of information through deep learning methods, such as using graph neural networks~\cite{papadimos2023flood}.

We also find papers describing ways of automatically aggregating and summarizing quantitative information, often for automatically creating maps that can complement--or ideally integrate--with those obtained by traditional processes.
Examples include, among many others, methods for accelerating damage assessment for creating a heatmap of potentially damaged areas in a region \cite{10.1126/sciadv.1500779.2016}, and for enhancing flood forecasts \cite{10.1109/wiiat50758.2020.00050.2020,10.1109/mis.2013.126.2013}.

These technological solutions tend to target the immediate response phase of a disaster, leaving open questions about how practitioners could be supported in the other phases of the EM lifecycle where they may use social media.
Our survey focuses on 
these questions, especially as we find that EM practitioners whose roles focus on emergency preparedness and recovery are increasingly using social media.

\subsection{Practitioners' View of Social Media}
As the next generation of EM practitioners becomes acclimated to 
their professional environment, it is worth noting that they were more widely introduced to technology growing up versus their veteran peers, and thus more likely to be social media-savvy users.  
As this demographic moves into different roles in EM agencies, they bring new skills 
and comfort with online spaces while also being aware of the public's social media adoption and expectations. Research indicates 
the public is increasingly and regularly consuming news on social media \cite{pew:2023aa}. In fact, a widely cited survey commissioned by the American Red Cross in 2010 \cite{redcross_2011}
found that 
80\% of the general population and 69\% of the online population surveyed expected 
national EM agencies 
to regularly monitor social media sites in order to respond promptly. A survey by Reuter and Spielhofer~\cite{reuter2017towards} in Europe in 2015 observed similar patterns regarding public expectations of 
response from EM agencies.     
Consequently, the literature around practitioners' use of social media across all phases of the EM lifecycle 
needs to be revisited and updated to account for 
new and evolving practices, expectations, and barriers. 

Prior research in crisis informatics often refers to surveys on social media usage by 
EM practitioners. 
For instance, 
Reuter et al.~\cite{reuter2016emergency} explored the attitudes of EM practitioners in Europe toward social media through an extensive survey in 2014, involving 
emergency service staff across 32 European countries. 
The authors 
found an overall positive attitude and the primary use to share information rather than for collecting information. 
Given the emergence of social media-savvy new practitioners, our survey and interview studies aimed to explore the patterns of social media usage for data collection practices of practitioners as well. 

Lorini et al.~\cite{lorini2021social} revisits the use of social media and describes the results of a workshop including both crisis informatics researchers and EM practitioners. 
On the one hand, this report indicates various aspects in which 
practitioners see value in social media: ``real-time and ongoing situational awareness; rapid insights in the immediate aftermath of disaster; integration with heterogeneous, multimodal data; and 
varying value across the EM lifecycle.'' 
On the other hand, they also point to various barriers to further adoption of social media. Information integrity issues, including verification and validation 
to prevent adverse consequences of misinformation, were the primary barriers expressed. 

The meta-analysis by Luna and Pennock \cite{luna2018social} cites various aspects that 
practitioners find valuable from social media, including increasing situational awareness, accelerating information diffusion, monitoring activities, and coordinating between 
stakeholders. 
At the same time, they describe both social and technical challenges. 
Among the social challenges, we again find information quality concerns, particularly the spread of false, incorrect, and outdated information, together with other concerns related to interoperability and the skewed demographics of social media users. 
Among the technical challenges, we find, among others, concerns related to scalability of data processing, 
dealing with unstructured data, the lack of standards and technical infrastructure, and  questions about data ownership and security. 

%
In general, an analysis of the literature shows that most of the practitioners' concerns have, to some extent, been considered by researchers who actively work on these issues, such as 
combating online misinformation and filtering information. 
However, practitioners might consider or perceive that the technical solutions to some of these problems have not been fully addressed 
to their satisfaction \cite{hiltz_exploring_2020}. For instance, one commonly perceived shortcoming is that current misinformation detection methods cannot detect all \inbrown{misleading messages}, or what 
EM practitioners would consider a sufficient fraction. 
Likewise, practitioners continue to encounter difficulties in moderating their spaces, addressing the continuing fragmentation in the information ecosystem, and responding to this fragmentation's differential impacts across demographic audiences, 
especially as information \& communication technologies continue to 
evolve rapidly. Further, most existing surveys are not conducted on an international scale to provide common patterns across  geographies and EM cultures, a gap that we aim to fill in this study. 
Together, these open concerns necessitate an updated investigation of the ways in which social media platforms integrate with practitioners' policies, procedures, and perspectives--this paper attempts to provide such an update.
\inblue{
\subsection{Traditional Communication View of Social Media}
The media plays a crucial role in shaping public perception and response in the aftermath of disasters. Vasterman et al.~\cite{Vasterman2005} describe how media coverage can inform the public, influence policy decisions, and mobilize resources for relief efforts. However, the authors also note that the media hype may distort the reality of the situation, create unnecessary panic, or shift focus away from essential recovery actions. The balance between trustworthiness and sensationalism significantly impacts how communities understand and react to disasters. 

Researchers highlight the critical role of social media in crisis communication for organizations ~\cite{ROSHAN2016350,hughes_evolving_2012,starbird_chatter_2010}. Social networks are pivotal in disseminating information rapidly and engaging directly with the public during crises \cite{olteanu_what_2015, williams_social_2018}. They provide a channel for real-time updates, which is crucial for managing public perception and providing accurate information promptly ~\cite{hughes_online_2014}. Social media enables organizations to monitor public sentiment and feedback, allowing 
more responsive and adaptive communication strategies ~\cite{st_denis_what_2020}. Moreover, these platforms facilitate two-way communication, fostering community and trust by enabling direct interaction between organizations and stakeholders ~\cite{ROSHAN2016350,hughes_evolving_2012}. This dynamic interaction helps mitigate misinformation and enhance the overall effectiveness of crisis management efforts ~\cite{arif_how_2016}.
Brandtzaeg et al.~\cite{Brandtzaeg2016} examined how 24 European journalists use and verify social media content. Journalists frequently monitor platforms like Twitter/X, Facebook, and YouTube, using tools like TweetDeck to manage information. They cautiously trust the content, often verifying it through multiple sources. Verification techniques include using Google Image Search and TinEye for photos and cross-referencing video locations with Google Maps. Despite the utility of social media, journalists emphasize the importance of rigorous verification to ensure accuracy.
In a 2016 study on politicians' crisis communication strategies, Jong et al.~\cite{Jong2016}, show how 
the 
effective use of social media \inbrown{by politicians} to quickly disseminate information, manage public perception, and engage directly with citizens helps in maintaining 
transparency and building 
trust during crises.

These studies collectively underscore how the integration of social media into traditional and political communication frameworks is enhancing engagement, transparency, and reliability during crises. Yet, there remain challenges. Given the concern about social media veracity from both practitioners and traditional media, as emerged from this study, there is a need to enhance the validation of user-generated data in terms of accuracy, and 
also processing time. 
}


%% file: 03-methods.tex
\section{Methods} \label{sec:methods}

This study uses a mixed-method approach to understand the changing practices of EM agencies. We first conducted a survey that asked close-form questions 
and also a few open-ended/ free-text 
questions 
(see details in the Appendix) 
and was distributed broadly. We followed this up with a smaller and more focused interview study. The interviews allowed us to ask more in-depth and open-ended questions than could be answered in the survey, and let us obtain a deeper understanding of 
EM practitioners'  
experiences and challenges. In summary, the interviews provided an in-depth examination of 
the practitioners' use of social media, while the survey provided a broader understanding of their usage. The methods used in these two studies are outlined below. 

\subsection{Survey}

We administered the survey online through Qualtrics and sought respondents' consent prior to their participation in the survey, as per the approved protocol for human subjects research study by the Institutional Review Board at the first author's institution. Participants were not required to complete all the questions in 
the survey and could decide to respond to either one or all study procedures, which is evident in the results later. 
We expected respondents to complete the survey in 15 minutes, maximum.  

\noindent \textit{Questions.} We created a set of questions as shown in Table~\ref{table:survey-questions_brief}, personalized for the U.S. and the European context, and geared towards participants according to their organizations' administrative levels 
(see detailed information in the Appendix, Table~\ref{table:survey-questions}). 
We also asked if they were available for a further study, which is how we recruited participants for 
follow-up interviews. 

\noindent \textit{Participants}. The survey participants 
were related to the 
EM profession, which we define as those who self-identify as currently working at an agency or organization involved with 
EM operations at any level in any EM phase, including 
local, regional, national, or international level. The participants were adults (age 18 and older) from diverse backgrounds. 
We did not follow any demographic inclusion/exclusion criteria except for asking about the participant's occupation background and work designation. For the scope of this study, we reached out to 
EM professionals working for agencies within the U.S. 
and European continent 
only.  

\begin{longtable}{
lp{0.9\textwidth}
} 
\caption{Survey questions. \inblue{(Full details of answer options for each question are provided in Appendix, Table~\ref{table:survey-questions}.)}
}\label{table:survey-questions_brief}\\ 
 \hline
 & \textbf{Questions}  
  \\ [0.5ex] 
 \hline\hline
 Q1.& Where do you work? 
  \\ 
 Q2.& What discipline does your agency represent? 
 \\ 
 Q3.& What is your position? (e.g., Firefighter, Fire Chief) 
 \\ 
 Q4.& Please describe your role/responsibility in one or two lines. 
 \\ 
 Q5.& At what level does your agency work for emergency management? 
 \\ 
 Q6.& How often does your agency use social media? (e.g., Daily, Weekly, Never)
 \\ 
 Q7.& How often do you use social media in a personal capacity (e.g., Daily, Weekly, Never)
 \\ 
 Q8.& What situational awareness value do you think social media could provide to your agency? 
 \\
 Q9.& What public communication value do you think social media could provide to your agency? 
 \\ 
 \hline 
& \textit{The following questions are only asked if the answer to Q6 is not `never':}  
 \\ 
 \hline 
 Q10.1. & What type of tools does your agency currently use to gather incident-specific information? 
 \\ 
 Q10.2.& What type of tools does your agency currently use to publicize incident-specific information? 
 \\ 
 Q10.3.& Using the above-listed tool(s), what does your agency use social media for? 
 \\ 
 Q10.4.& What do you consider the most critical barrier(s) to emergency management agencies not using social media in an emergency operations center environment? 
 \\ 
 \hline
\end{longtable} 

\noindent \textit{Dissemination.} We sent the link to the online survey 
to different networks of 
EM practitioners  
as well as national and international mailing lists that included the 10 U.S. Regions of the International Association of Emergency Managers (IAEM),  the European Emergency Number Association (EENA), and Emergency Services Staff Network (ESSN). Additionally, the survey was shared via LinkedIn with the International Association of Chiefs of Police (IACP), the International Association of Fire Chiefs (IAFC), the International Association of Fire Fighters (IAFF), and the International Public Safety Association (IPSA).  The data collection period covered six months, March 15 to September 15, 2022.   
We also provided anonymity to the survey respondents by ensuring that their responses would not be linked to their identities except for those who volunteered to further participate in the future research activities of this project. Specifically, we collected e-mail addresses but kept them separate from survey responses for those who volunteered to participate in further studies--such as an interview or a focus group--or who would like to receive the publications resulting from the data analysis of the survey responses. 

\noindent \textit{Analysis}. To analyze the survey responses, we conducted both quantitative and qualitative analysis to discover patterns from closed and open-ended/ free-text questions. 
For the quantitative analysis, we analyzed the responses using Excel software by first exploring basic descriptive statistics for a question (e.g., frequency by location, the discipline of participants' agency such as Fire Service, the agency's level such as county/local, etc.). Further, we employed cross-question or cross-tabulation analysis to find patterns relating to participants from different agency levels and their perceptions such as the value of social media for situational awareness. 
For the qualitative analysis of free-text survey questions, we used an inductive approach of grounded theory for coding to derive categories from the free-text answers ~\cite{glaser_discovery_1967}, e.g., the type of position and role. Two authors coded each response to free-text questions openly and divided them into one or multiple categories to derive themes based on the prior knowledge from the literature review. There were a few responses in the non-English language, which were manually translated into English using Google Translate.

\subsection{Interviews}
Following the survey, to better understand the experiences of EM practitioners, 
we conducted interviews with 11 participants 
from different locations and areas of the EM work functions 
(see Table~\ref{tab:interview_participants}). 
We recruited 
interviewees from the pool of survey participants who volunteered to participate in a follow-up interview for this project. This voluntary sample was not intended to be representative of all EM practitioners, but rather it allowed us to 
complement 
the breadth of the survey results with a more in-depth understanding of our survey participants' experiences. 
We conducted interviews remotely via Zoom, with the 
goal of a one-hour interview; we saved recordings and transcriptions for later analysis. 
As with the survey, we obtained approval for the human subjects research study for this research from the first author's 
institution 
and followed all ethical protocols. 

\begin{table}[t]
\centering
\caption{Summary of the interview participants. \inblue{(Full details of interview questions in Appendix, Table~\ref{table:interview-questions}.)}}
\label{tab:interview_participants}
\begin{tabular}{l p{2in} l p{1.5in}}  
 \hline 
 \textbf{Participant}  & \textbf{Organization} & \textbf{Agency Type} & \textbf{Primary Tasks} \\ [0.5ex] 
 \hline\hline
 P1 & Government & County & Public 
 communication 
 \\ 
 %
 P2 & Community Organization & County & Information filtering \\
 %
 P3 & Emergency Management & County & Preparedness and training \\
 %
 P4 & Emergency Management & County & Planning section    \\
 %
 P5 & Emergency Management & County & Overall emergency management \\ 
 %
 P6 & Transportation Department and Community Organization & State &  Information collection for  
 operations   \\ 
 %
 P7 & Meteorology   & National & Weather related intelligence   \\ 
 %
 P8 & Community Organization & National & Monitoring and information filtering \\ 
 %
 P9 & Community Organization & National  &  Monitoring and information filtering \\ 
 %
 P10 & emergency response center & Continent  & Geospatial intelligence   \\ 
 %
 P11 & emergency response center & Continent  & Public communication  \\ 
 \hline
\end{tabular}
\end{table}

\noindent \textit{Questions.} 
Interview questions (see details in the Appendix, Table~\ref{table:interview-questions}) asked about individuals' experiences in disaster events, 
and their use of social media for gathering information and communicating with the public. We asked about their situational awareness needs and how they had used social media to meet these needs. The interviews also examined inter-agency social media communication, integration within agency operations and communication plans, and the presence of Standard Operating Procedures (SOPs) for social media during disasters.

\noindent \textit{Analysis.} 
To analyze the interviews, we used an inductive, grounded theory approach ~\cite{glaser_discovery_1967}. Three of the authors read all of the interview transcripts and extracted excerpts from the interviews that answered each of the questions we asked, as well as passages that were particularly salient or insightful. Then these excerpts were 
open-coded ~\cite{charmaz_constructing_2006}. The three authors met several times to iteratively consolidate, refine, and discuss their codes until \inbrown{a} consensus was reached on our code applications. During these sessions, 
we identified common themes from the codes using thematic analysis \cite{braun_thematic_2012}. These themes were then refined through bi-weekly meetings with the whole project team. We found that our codes broadly fell into four types of communication practices informed by the literature ~\cite{10.1016/j.giq.2012.12.004,10.1016/j.giq.2013.05.015,https://doi.org/10.1111/1468-5973.12119, 10.1016/j.giq.2017.10.003}: dissemination, collection, engagement/mobilization, and cooperation. We used these themes to derive a communication model (described in more detail below) for thinking about the impact that social media has and continues to have on emergency management. 

%% file: 04-results-survey.tex
\section{Findings from the Survey} \label{sec:findings-survey}


We collected a total of 160 
survey responses. 
We 
noticed a handful of responses from non-EM professionals that were discarded, resulting in 153 responses for analysis. Below, we 
summarize responses to the 
survey questions listed in Table~\ref{table:survey-questions} in the Appendix, highlighting the common patterns in terms of social media usage across the U.S. and European countries. We also note that given the call for voluntary participation in the survey, and that not all questions were mandatory, not all participants answered every question. 
Despite this limitation, we have, at minimum, 83  
responses per closed-form question, and our qualitative analysis suggests sufficient saturation of responses. That said, an anonymized version of this dataset for further research exploration, as approved by the Ethics/Institutional Review Board, is available upon request to authors. 

To enhance the clarity of our cross-question analysis, we manually categorized the working level of a participant's agency into two broad categories based on the response to question Q5: \textit{local} (if the response included  `County/Local' or 
'Municipal/Town/Local') and \textit{non-local}. This decision was driven by the recognition that the responsibilities of agencies working at the \textit{local} level can vary significantly from others. 
This approach allows us to focus on the key differences in the value of social media for situational awareness and public communication to local versus non-local agencies, providing more  
meaningful insights. 
Non-local agencies, including national, state, and regional authorities, operate on a broader scale, covering larger geographic areas, and providing assistance, resources, and coordination for multiple local agencies. 
During major disasters, these higher-level agencies coordinate with local governments to supplement and enhance their response capabilities. They work to ensure effective utilization of resources, always recognizing that the primary responsibility remains with local authorities, as all disasters are fundamentally local in nature. 
Local agencies handle immediate response, while non-local entities provide broader support and coordination. Thus, such a distinction may also explain the differing values and answers within the survey questions when comparing local to non-local agencies, as their perspectives and responsibilities differ significantly in terms of scale and scope of operations. 

\noindent \textbf{
\textit{
\inblue{1. Representative Agency Location, Discipline, and Level}.}}  
We received a higher level of responses \inblue{(Q1)} from the U.S. (47\%) and Portugal (36\%),  
followed by other nations \inblue{(17\%)} including Croatia, France, Germany, Greece, Italy, Netherlands, and Spain.   %
%
Approximately \inblue{half 
of the responses (Q2) were from agencies with a mandate that can be categorized as}  
civil protection and emergency management,  
14\% health and medical services, 11\%  fire services, 6\% telecommunications, 5\% non-governmental organizations, 2\% police and security services, 1\% transportation, and 9\% others. 
%
The participants were involved in different positions \inblue{(Q3)} in agencies  
 relevant to 
 EM profession.  
These roles included Emergency Manager, Emergency Services Coordinator, Public Information Officer, EM Planner, Fire Chief, Firefighter, Emergency Management Specialist, Battalion Chief, 
Emergency Preparedness Manager, etc. 
%
\inblue{Over 60\%} of the responses  
regarding the roles indicated that 
practitioners take on \emph{multiple roles} within their organizations. This aligns with previous research \cite{hughes_evolving_2012,hiltz_exploring_2020}, highlighting a common barrier faced by emergency managers when incorporating social media into 
their environment. 
Such a large number of participants reporting multiple roles also suggests that 
the agencies are often 
short-staffed, which could impact the successful integration of social media \inblue{in their practice}.   
\inblue{This insight informed our interviews to explore the types of responsibilities associated with multiple roles, when collecting information for situational awareness or communicating to the public.} 

\inblue{Regarding the level (local/non-local) of a participant's agency, the responses show that 
approximately 
51\% 
\inblue{(Q5)} 
indicated affiliation with agencies working at the local level, including   
county, municipal, or town-level agencies. The local level is the key to successful EM operations for an event, 
as all disasters begin at the local level and then their response expand to the state and national level\inbrown{s} when the scope, magnitude, and complexity increase.} 
%

In addition, the responses represented work responsibilities \inblue{(Q4)} focused on distinct phases of the EM lifecycle. The highest 
percentage of answers fell into 
the category of mitigation and preparedness, followed by response. 
This observation challenges the expectation, \inblue{as per the literature review,} that social media adoption and use predominantly occur 
during the response phase, as 
often emphasized by the considerable focus of crisis informatics research on disaster response \cite{10.1126/science.aag2579.2016ogo}. This \inblue{indicates an opportunity for researchers} to direct attention beyond the response phase.

\begin{figure} 
\centering
\begin{subfigure}{0.62\textwidth} 
\includegraphics[width=\textwidth]{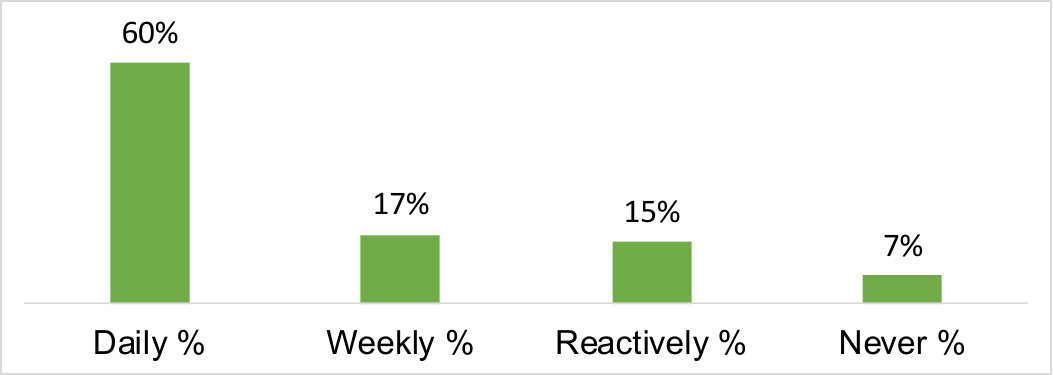}
\caption{Frequency of using social media at the agencies of participants (Q6).} 
\label{fig:agency_frequency}
\end{subfigure}
\vspace{1em} 

\begin{subfigure}{0.62\textwidth}
\includegraphics[width=\textwidth]{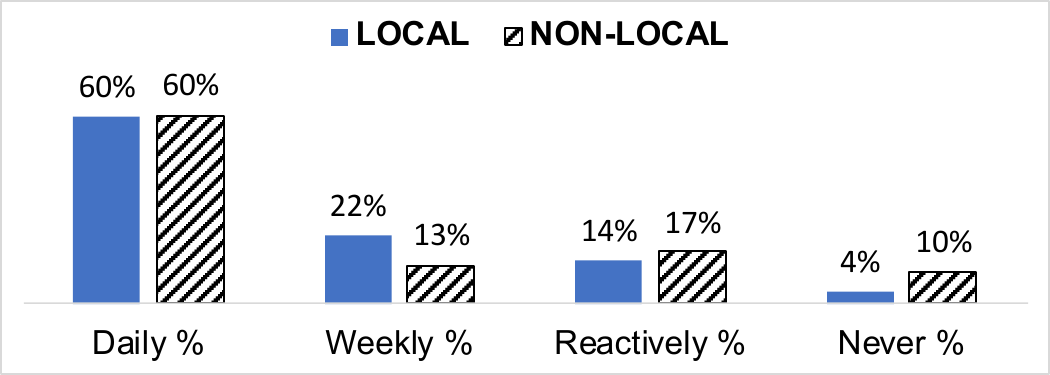}
    \caption{Frequency of using social media at different levels of agencies (Q5 \& Q6).}
\label{fig:agency_frequency_for_social}
\end{subfigure} 
\vspace{1em}

\begin{subfigure}{0.62\textwidth}
\includegraphics[width=\textwidth]{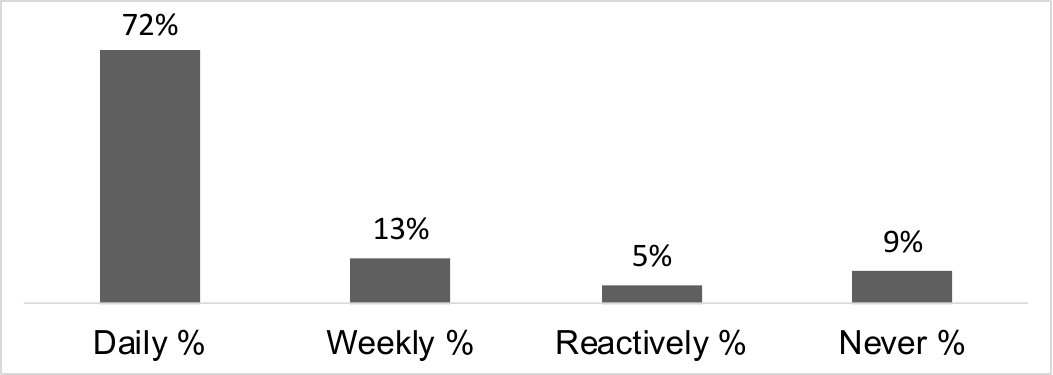}
    \caption{Frequency of using social media personally (Q7).}
\label{fig:personal_frequency}
\end{subfigure}
\caption{\inblue{Frequency of social media use in professional and personal capacity. (\emph{Reactively} implies the use only when an incident occurs.)}} 
\label{fig:agency_use_social}
\end{figure}

\vspace{1mm}
\noindent \textbf{
\textit{\inblue{2. Frequency of Social Media Use} in Personal and Professional capacity.}} 
A high percentage of survey participants \inblue{(Q6)} reported daily or weekly use of social media  
at their agencies \inblue{(over 75\% overall (Figure~\ref{fig:agency_frequency}), and over 80\% at local and over 70\% at non-local agencies (Figure~\ref{fig:agency_frequency_for_social})), indicating that it is becoming a common 
practice. About 10\% participants from non-local agencies indicated that they never use social media professionally, compared to 
only 4\% of participants from local agencies. Moreover, 63\% participants from the U.S. and 57\% from 
European countries indicated daily use of social media at their agencies.} 
This could be interpreted as a sign that EM agencies recognize the increasing necessity of social media, whether for attaining situational awareness, engaging with the public, or other purposes. 
%
\inblue{In personal capacity (Figure~\ref{fig:personal_frequency}), nearly 85\%} of the participants \inblue{(Q7)} indicated that they use social media either weekly or daily, which gives credibility to them for being \inblue{familiar enough with social media to use it professionally} in their agencies. %

\inblue{
The extent of using social media in both personal and professional capacities, 
coupled with the insights 
about the diversity of roles of participants in the survey provides credible support to conclude how social media has become a non-traditional, supplementary source of information for practitioners across different agency levels and regions.} 

\begin{figure}[b]
\centering
\begin{subfigure}{0.92\textwidth} 
\includegraphics[width=\textwidth]{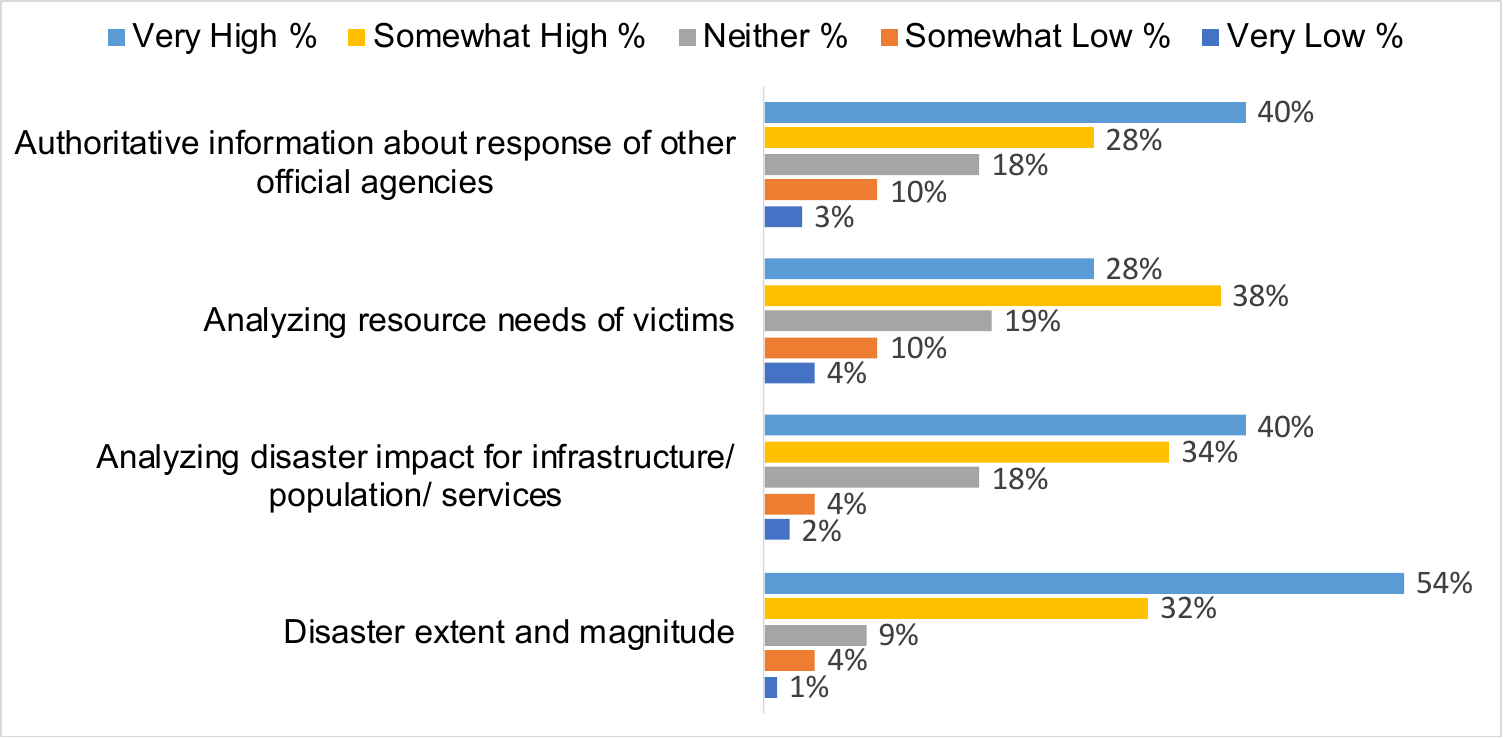}
    \caption{For Situational Awareness \inblue{(Q8)}} 
\label{fig:perceived_situational}
\end{subfigure} 
\vspace{1em}

\begin{subfigure}{0.92\textwidth}
\includegraphics[width=\textwidth]{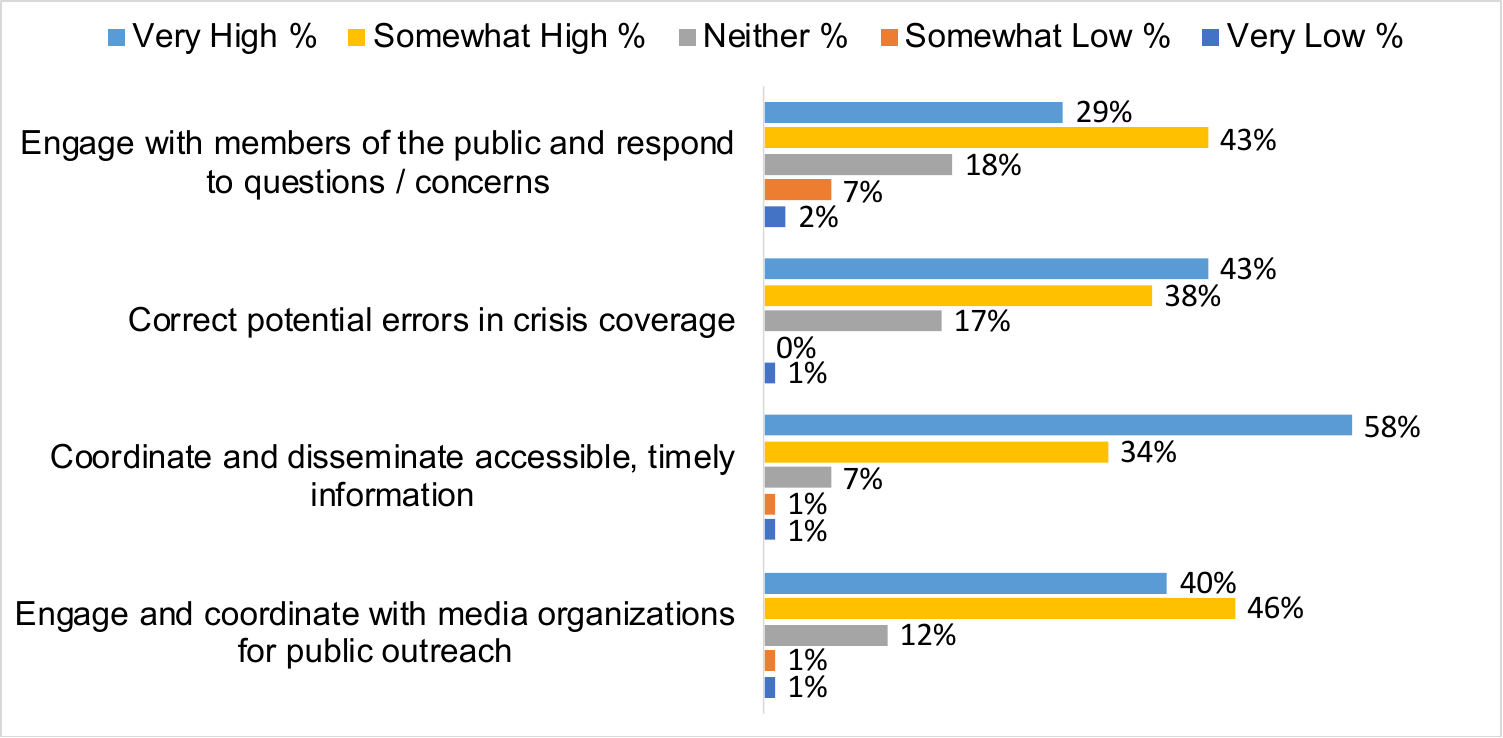}
    \caption{For Public Communication \inblue{(Q9)}} 
\label{fig:perceived_communication}
\end{subfigure} 
\caption{\inblue{Perceived value of social media} by survey participants in EM profession.} 
\label{fig:perceived_value}
\end{figure}

\vspace{1mm}
\noindent \textbf{
\textit{\inblue{3. Perceived Value} of Social Media}}.  
\inblue{As shown in Figure~\ref{fig:perceived_situational}, in three of the four factors 
to gauge the perceived value of social media for} situational awareness \inblue{(Q8)}, 
the majority of the participants selected a \textit{very high} value. 
The one situational awareness factor 
that did not receive a \textit{very high} value was analyzing the resource needs of the victims. This may be due to the fact that sheltering, feeding, and tending to the needs of victims is a specific function in disaster response, whereas the other high-value factors (disaster extent and magnitude, analyzing disaster impact, and attaining information about other agencies) are broad and can be applied to numerous functions. 

\inblue{Similarly as shown in  Figure~\ref{fig:perceived_communication}, 
the perceived value of social media for public communication \inblue{(Q9)} 
 is considered positively by a majority. In particular,} a high percentage of responses (over 90\%) indicated that social media could provide \textit{somewhat high} or \textit{very high value} for coordinating and disseminating accessible, timely information. Of these responses, 
 58\% chose a\textit{very high} versus a \textit{somewhat high} value. This is a strong indicator of the positive mark social media has made within EM agencies, \inblue{supporting the claims from prior surveys in crisis informatics that the use of social media has evolved to become a mainstream,  traditional practice in EM.}  

\begin{figure}
\centering
\begin{subfigure}{0.78\textwidth}
    \includegraphics[width=\textwidth]{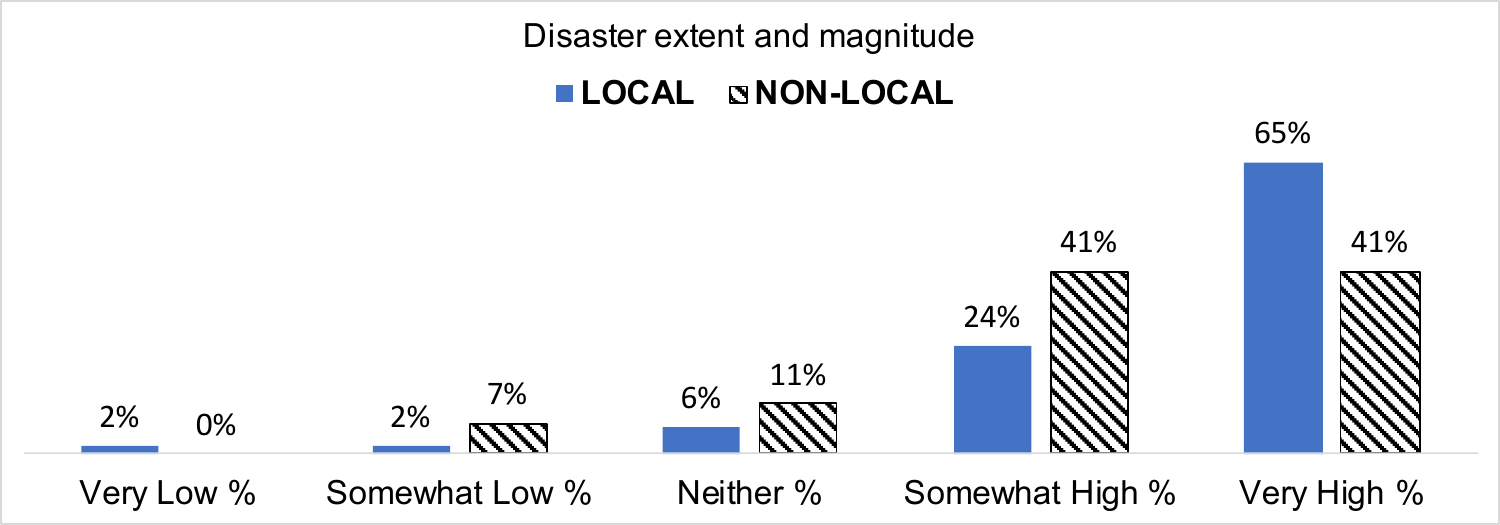}
    \caption{}
    \label{fig:q8_1}
\end{subfigure}
\hfill
\begin{subfigure}{0.78\textwidth}
    \includegraphics[width=\textwidth]{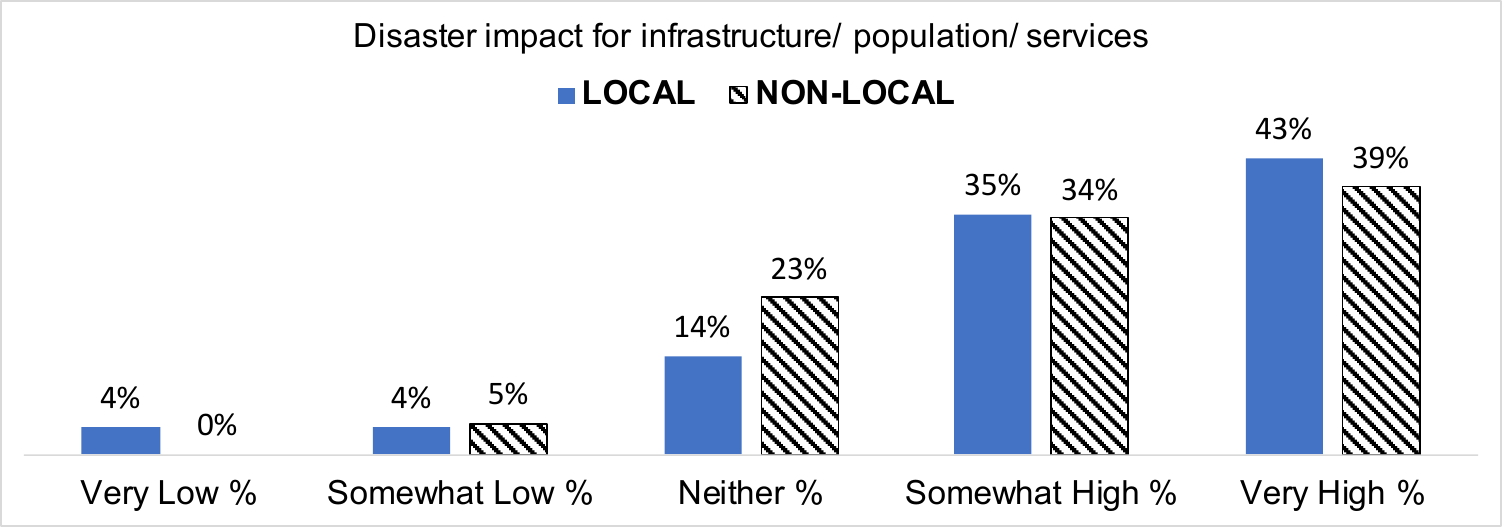}
    \caption{}
    \label{fig:q8_2}
\end{subfigure} 

\begin{subfigure}{0.78\textwidth}
\includegraphics[width=\textwidth]{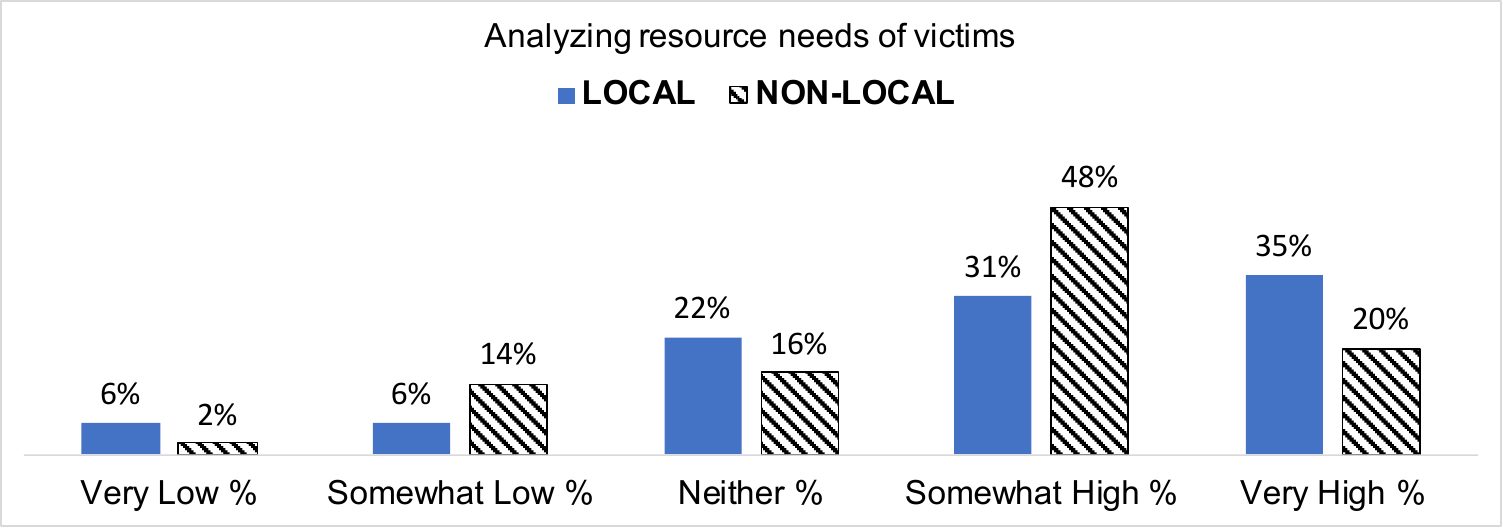}
    \caption{}
    \label{fig:q8_3}
\end{subfigure}
\hfill
\begin{subfigure}{0.78\textwidth}
    \includegraphics[width=\textwidth]{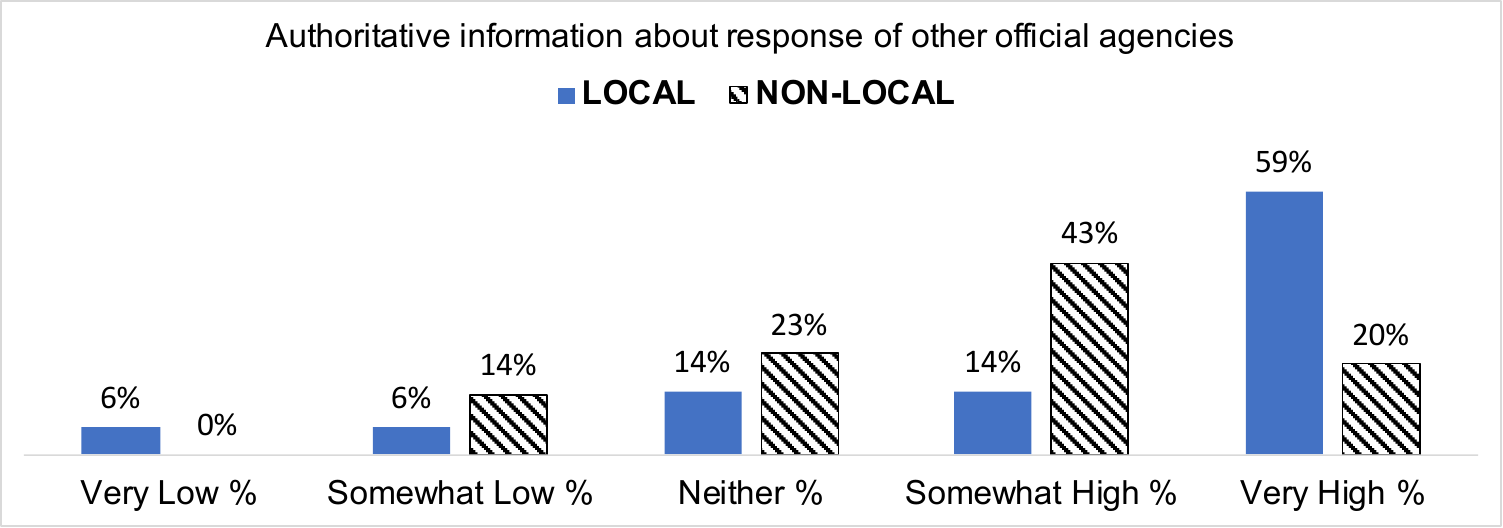}
    \caption{}
    \label{fig:q8_4}
\end{subfigure} 
\caption{\inblue{Perceived value of social media for situational awareness by survey participants across agency levels}.} 
\label{fig:q8_SA}
\end{figure}

\begin{figure}
\centering
\begin{subfigure}{0.78\textwidth}  \includegraphics[width=\textwidth]{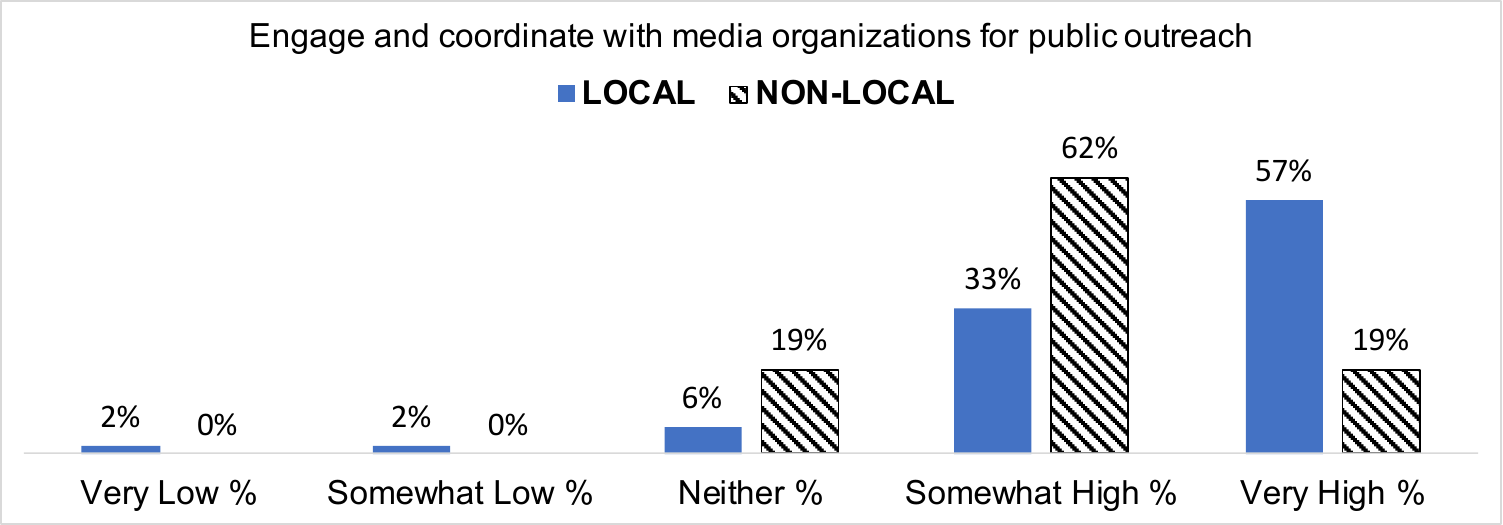}
    \caption{}
    \label{fig:q9_1}
\end{subfigure} 

\begin{subfigure}{0.78\textwidth}  \includegraphics[width=\textwidth]{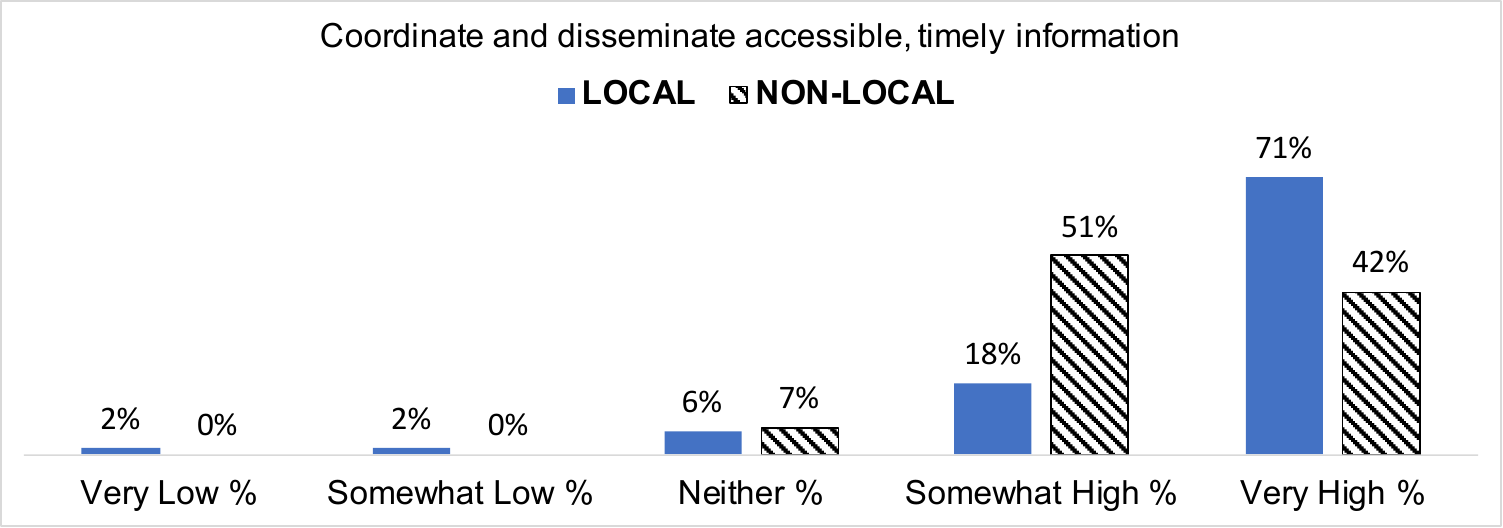}
    \caption{}
    \label{fig:q9_2}
\end{subfigure} 

\begin{subfigure}{0.78\textwidth}
\includegraphics[width=\textwidth]{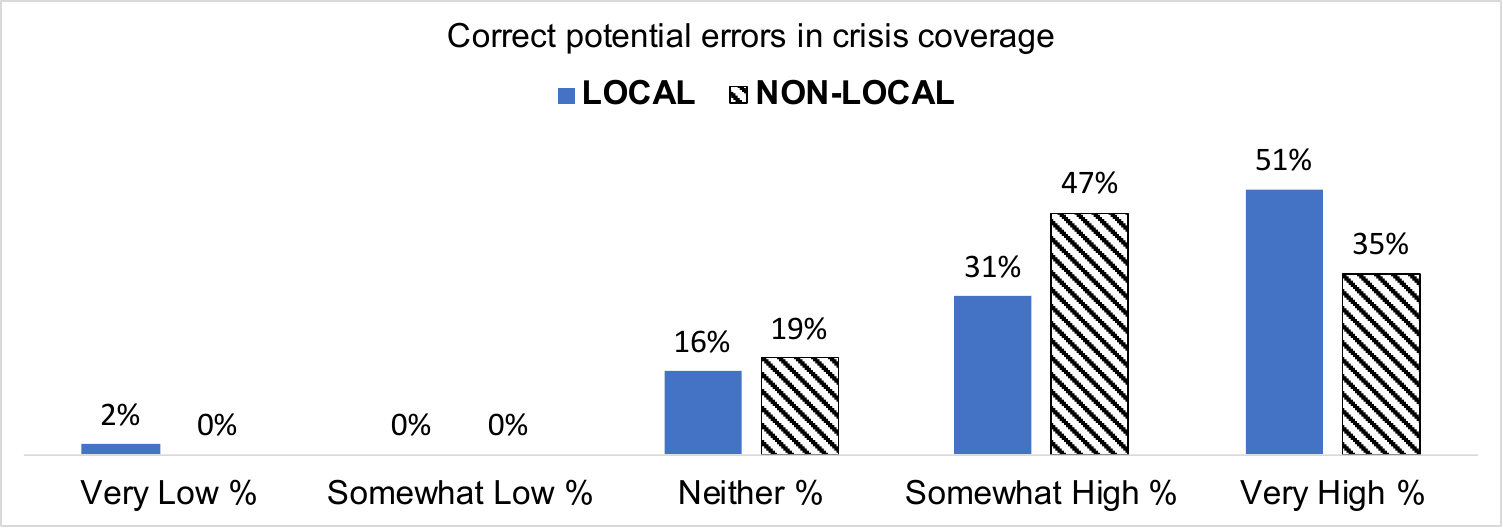}
    \caption{}
    \label{fig:q9_3}
\end{subfigure}

\begin{subfigure}{0.78\textwidth} 
\includegraphics[width=\textwidth]{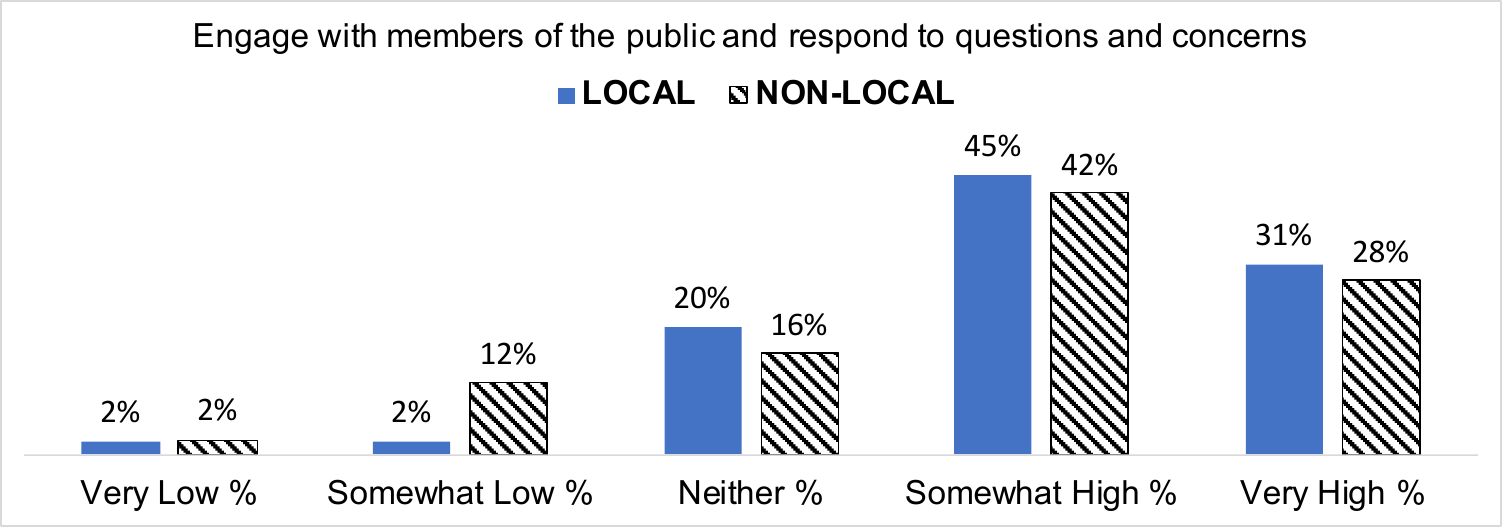}
    \caption{}
    \label{fig:q9_4}
\end{subfigure} 
\caption{\inblue{Perceived value of social media for public communication by survey participants across  agency levels}.} 
\label{fig:q9_PC}
\end{figure}

\inblue{Furthermore, Figure~\ref{fig:q8_SA} and \ref{fig:q9_PC} show results by agency level. 
Responses affiliated with local level agencies indicate a \textit{very high} value of social media for analyzing disaster extent and magnitude, as opposed to non-local agencies. Perhaps due to the potential to acquire actionable information for local response, 
especially when considering their daily usage practice as observed in Figure~\ref{fig:agency_frequency_for_social}. This trend is further evident in the perceived value \inbrown{of social media} for public communication, especially for engaging with media organizations and correcting errors in crisis coverage.}

\vspace{1mm}
\noindent \textbf{
\textit{ \inblue{4. Common Tools and Purpose of Use} 
by Agencies}}. 
Over 30\% of the responses \inblue{(Q10.1) 
mentioned tools for gathering incident information and} community engagement on social media like 
Hootsuite or TweetDeck (or both). Examples of the wide array of different tools 
listed by the participants 
include operational management tools like WebEOC, data analytics tools like DataMinr and Sprout Social, and communication tools like Facebook Messenger and WhatsApp. Some participants' responses reported using general 
social media platforms like Twitter/X and Facebook as their tool and also, data feeding tools like Google Forms in their current usage. 
%
%
Approximately 40\% participants \inblue{(Q10.2)} 
listed Facebook or Twitter/X as social media tools that their agencies use in publicizing incident-specific information. 
However, many of the other responses were ambiguous (e.g., ``social media'') and thus,  
this percentage might be greater. 
%
%
%
\inblue{For the purpose of use (Q10.3), it is interesting to note that} the purpose of tools for continuous monitoring received a low percentage (\inblue{less than 20\%})  
as well as complementing previously known information.  
\inblue{It may be due to the additional resources required by} 
both of these tasks, 
conflicting with the limited human resources available in the 
EM agencies. 
A promising insight 
is that a 
substantial percentage of participants (24\%) listed incident-based monitoring as a function their agency uses with existing tools. 
Further, there were around 6\% 
participants who answered the choice of `Other', out of which 
58\%  
stated that they use social media in some form of communication capability.

\vspace{1mm}
\noindent \textbf{
\textit{5. \inblue{Barriers to Social Media Integration} within Emergency Operations}}.  
36\% of the responses \inblue{(Q10.4)} selected lack of human resources and/or management support as the most critical barrier to social media use in emergency operations centers.  Unverified information or sources received the second highest number 
of responses (29\%). This insight informed our questions for interviews regarding public communication. 
10\% 
participants selected the choice of `Other', 
out of which ~40\% of responses associated 
the lack of training to be 
the barrier. 
\inblue{This may point to the need 
for designing a solution that requires minimal intellectual knowledge of 
technology, thus avoiding the ``training'' barrier. It may also demonstrate the need for better institutional education regarding policies and procedures for social media use.} Additional 
barriers of interest included 
lack of policies, decision-making, connectivity/ technology in general, funding, and 
subjectiveness and credibility of social media information. 

\inblue{We use the above insights, together with the qualitative analysis of the next section, in developing our rationale for design needs, ranging from policies and procedures to technology, in Section \ref{sec:discussion}}.

%% file: 05-results-qual-analysis.tex

\section{Findings from Qualitative Analysis} \label{sec:findings-focus}

\begin{figure}[htbp]
\begin{center}
\includegraphics[width=0.67\textwidth]{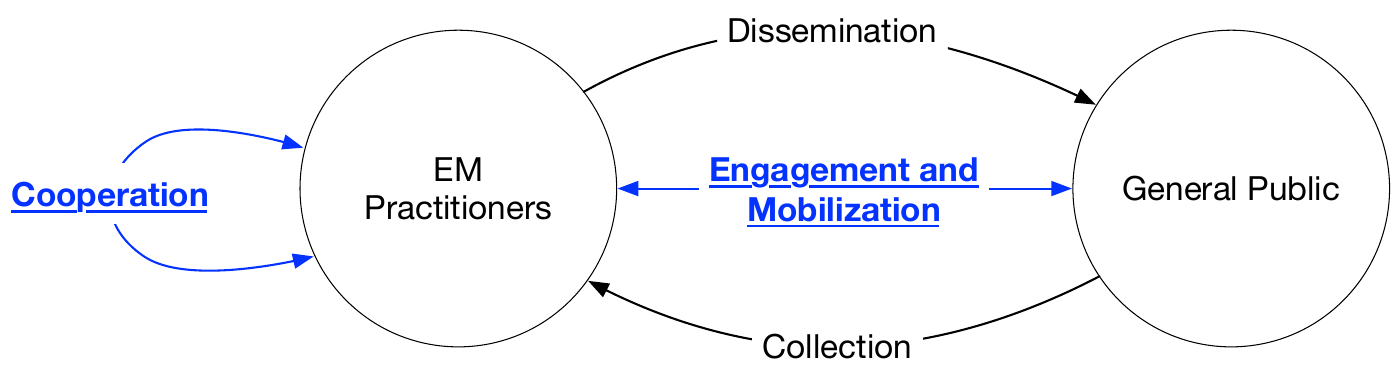}
\caption{Extended model of the communication processes among diverse practitioners and the public. Interviews reveal new methods, challenges, and approaches across these processes, especially with respect to cooperation and engagement/mobilization, where the crisis informatics literature has relatively little focus.}%
\label{fig:themechannels}
\end{center}
\end{figure}

Moving to the more in-depth interviews, we organize our findings around the four \inblue{communication processes shown in Figure \ref{fig:themechannels}: dissemination, collection, engagement/mobilization, and cooperation.}
\inblue{This extended model of communication processes 
merges and extends those presented in Mergel \cite{10.1016/j.giq.2012.12.004,10.1016/j.giq.2013.05.015},  Wukich \cite{https://doi.org/10.1111/1468-5973.12119}, and DePaula et al. \cite{10.1016/j.giq.2017.10.003}.}
\inblue{While descriptions of these communication processes are not new, research across them has mainly come from general studies of governmental communication and has used a variety of languages 
to describe their core concepts--Table \ref{tab:framework_history} helps unify this varied language.} 
\inblue{Despite research from governmental communications, much of the crisis informatics literature has focused mainly on the one-way/push-pull \emph{dissemination} and \emph{collection} processes, often to the detriment of other pathways.} 
\inblue{This section, therefore, focuses on the unique needs of EM practitioners \emph{across this extended model}, 
where preparedness, rapid response, and cooperation with volunteer groups are more salient, and EM practitioners increasingly engage in more two-way communication and coordination.} 
\inblue{From our interviews, we find that practitioners are equally focused on and face barriers regarding actively engaging with/mobilizing} the general public and  \inblue{regarding how they cooperate and coordinate across agencies and volunteer groups}. 

\begin{table}[bpth]
\centering
\caption{\inblue{An integrated communication framework for EM practitioners on social media}.}
\label{tab:framework_history}
\begin{tabular}{p{1in} p{1.75in} p{1in} p{1.25in}}  
 \hline 
 \textbf{Communication Process}  & \textbf{Definition} & \textbf{Related Terms} & \textbf{Prior Studies} \\ [0.5ex] 
 \hline\hline
\inblue{Dissemination} & Unidirectional pushing of information from EM practitioners to the public, primarily to alert, enhance awareness, or otherwise disseminate useful information. & broadcast, information provision, one-way, push, representation, transparency & \citet{10.1016/j.giq.2012.12.004,10.1016/j.giq.2013.05.015,10.1016/j.giq.2017.10.003,10.1016/j.giq.2020.101486,10.1016/j.giq.2020.101539,10.1145/2675133.2675242,10.1177/1461444816645652,https://doi.org/10.1111/1468-5973.12119} \\ 

\inblue{Collection} & Unidirectional consumption of content provided by the public via the data collection processes of EM practitioners, often to facilitate their 
situational awareness and reporting. & pull, participation, input-seeking, engagement$^\star$ & \citet{10.1016/j.giq.2012.12.004,10.1016/j.giq.2013.05.015,10.1016/j.giq.2017.10.003} \\ 

\inblue{Engagement and Mobilization} & Bidirectional, rapid, and synchronous engagements between EM practitioners and the general public, often with the goal of quickly soliciting actions or responding to requests or reports from the public, and with an element of turn-taking between a practitioner and member of the public.  & collaboration, crowdsourcing, networking, online dialog, responsiveness, two-way communication, two-way dialogue, interactivity & \citet{10.1002/pa.1967, 10.1016/j.giq.2012.12.004, 10.1016/j.giq.2013.05.015, 10.1016/j.giq.2017.10.003, 10.1016/j.giq.2019.03.001, 10.1016/j.giq.2020.101486, 10.1016/j.giq.2020.101539, 10.1177/1461444816645652} \\ 

\inblue{Cooperation} & Bidirectional engagements \emph{between EM practitioner} organizations, both government- and private citizen-/volunteer-run, to coordinate responses and share information \emph{as peers} in the EM practitioner space. & coordination, cross-agency dissemination & \citet{10.1016/j.giq.2020.101539,10.1002/pa.1967} \\ 

 \hline
 \multicolumn{4}{l}{\makecell[l]{
 \footnotesize $^\star$ This ``engagement'' means non-responsive collection of citizens' responses to government agency posts;\\ \footnotesize no bidirectional engagement is implied.}}
\end{tabular}
\end{table}

Regarding (1) \textit{dissemination}, we identify new concerns, challenges, and approaches around the sharing of information with the general public--e.g., disseminating messages across a highly fractured social media ecosystem, producing visual media rather than simply text, to targeting second-order social connections. 
Beyond practitioners' dissemination practices, 
a much-studied use case for social media spaces has been for practitioners to collect data from these spaces, much of which is for situational awareness~\cite{vieweg_microblogging_2010, zade_situational_2018}. Our new findings, however, highlight new approaches to (2)  \textit{collection} that practitioners have begun to employ 
to address both new and long-standing concerns with social media data.   
%
Furthermore, rather than seeing a clear delineation between practitioners simply disseminating and collecting information to/from the public, we find new processes around (3) \textit{engagement and mobilization}, where practitioners actively solicit information, request action, and respond to requests from the public before, during, and after an event. 
Lastly, we find these online social spaces are not just spaces for practitioners to communicate with the general public. These spaces also provide new avenues for (4) \textit{cooperation} among and across practitioner organizations.

\subsection{Dissemination Practices}
\label{sec:findings-focus-dissemination}

\inblue{Public communication was the most frequently cited purpose for using social media in our survey, encompassing one-third of the responses. Through 
interviews, we gathered insights about various aspects of interviewees' dissemination practices.}

\paragraph{Social Media as Traditional Media.} 
EM practitioners continue to place significant reliance on conventional media outlets, including ``television and radio stations'' [P2]  
and ``press releases'' [P11]. 
A participant emphasized it as, [P1]:  
``Traditional media has certainly been a key way [we communicate with the public]. We've reached out with news conferences, news releases, [and] interviews.'' 
At the same time, however, the distinction between traditional media and new social media has become less clear. 
Social media sources, once considered non-traditional owing to their novelty, have now become integral and mainstream communication tools. 
One participant expressed this shift succinctly, stating, 
[P11]  
``I'm not sure any more what non-traditional really is. Because, for me, social media has become traditional and mainstream.''  
This blending is clear in responses from 
[P10] 
and [P5],  
where their dedicated communication units manage Twitter/X accounts, and [P11]'s  
comment that social media is ``just another announcement similar to [a] press release.''

\paragraph{Social Media and Emergency Preparedness.} This integration of traditional and social media also means that social media has become a resource during so-called ``blue-sky days'', when no emergency is happening, and practitioners can share messaging on emergency preparedness. 

\begin{displayquote}
\emph{[P4]}  
During blue sky days we do a lot of positive communication. We have a newsletter that comes out on a monthly basis. We do regular Facebook and Twitter posts. We make it a point to do at least one post a day. We do a lot of campaigns, so depending on if there is a weather event or some kind of other emergency that's happening around the world, we would capitalize on that.''
\end{displayquote}

These blue-sky days provide a particularly impactful channel when a high-impact disaster has occurred elsewhere, such as with the 6 February 2023 earthquake in Turkey \cite{turkey_earthquake_wiki}.  
The collective awareness of the public of this event allowed practitioners to leverage it into emergency preparedness messages for similar events: 
[P4]  
``After the shock [has] worn off about the earthquake in Turkey, we took that opportunity to... get back with our residents and make sure that they know what to do if an earthquake would ever hit here.''

\paragraph{Access to New Audiences.} For these blue-sky days, social media has also offered new audiences to practitioners. 
[P11]  
describes how Instagram is particularly useful for reaching younger audiences. 
These new audiences are not limited to just more diverse demographics: [P4]  
points to the use of social media 
for reaching indirect but important audiences, such as the non-local families 
of elderly individuals or vulnerable populations. 

\begin{displayquote}
\emph{[P4]}  
Social media for us is unique because we might have people all around the country and the world that are following us, because they might have family here. So, using the messages, we are still getting to residents, whether it is directly or indirectly from their family members or friends who might see things that we post. 
\end{displayquote}

\paragraph{Near Real-Time Dissemination.} A critical point mentioned about these social media channels, in contrast to traditional media, was the speed with which practitioners could respond to and capitalize on foreign events and local emergencies.
For example, [P9]  
describes how their main channels of communication were Instagram and Twitter/X, where they ``tweeted a lot because [it's] real time... It's a nice option because it allows you to send a sequence of messages in real time.'' 
This real-time affordance is not universal across social media platforms, however; [P9]  
 makes specific mention of Facebook, where real-time communication is difficult. 
[P11]  
makes a similar comment about 
differences in platforms 
and the importance of speed: ``We first publish things on Twitter. We do try to publish something on Facebook and Instagram.''

\paragraph{Dissemination in a Fragmented Space.} The challenge of managing information dissemination during a disaster has become increasingly complex due to the proliferation of social media platforms. 
Practitioners face challenges with fragmentation, as there are now too many platforms to both push information to and monitor effectively. 
Adding another channel adds new challenges, like managing yet another source of information: [P11]  
``We have a very small team. We are very overstretched, and recently there are more and more disasters, and wildfires are becoming more intense.'' 
This dilemma is compounded by the absence of a guarantee that the desired audience is being reached. 
As [P5]  
notes, ``When it first came out, you just had 2 or 3 platforms. Now you have so many platforms out there, and unfortunately, it gets to the point where people's viewpoint determines which platform they use. How do you reach all these people on these different platforms?'' 
This fragmentation poses a significant hurdle for emergency managers seeking to reach diverse populations during a crisis. 
[P1] 
highlights this concern as well, describing the social media space as ``becoming more fragmented, and it's hard to reach a massive people.'' 
Additionally, there is a concern about the perception of favoritism if an organization maintains an account on one platform over another. 
The evolving landscape of social media has made the task of communicating information during emergencies considerably more challenging. As [P3] 
notes, ``If everybody leaves Facebook and just goes to TikTok, and we have no way to reach them unless we find some kind of influencer that would push our information out to them''.

\subsection{Collection Practices}
\label{sec:findings-focus-collection}


New platforms, new affordances, and new procedures for sharing information have given rise to multiple changes in how practitioners disseminate information, as shown above, but when it comes to \emph{collecting} information from social media spaces, many of the same concerns outlined elsewhere (e.g., \cite{lorini2021social, hiltz_exploring_2020}) remain.
\inblue{Many survey respondents indicated that they did not use social media for continuous monitoring, but instead used it as a way of collecting complementary data, in an incident/event-specific manner.} 
Unsurprisingly, concerns around information quality, location, completeness, and timeliness were discussed by numerous interviewees. 

\paragraph{Questions of Accuracy and Location.}  
[P3], [P4], [P7], [P8], and [P9],  
all point to concerns around the geolocation of information as a critical problem: [P9] 
``Geolocation is one of the most difficult things to do through social media,'' and [P8] 
``The challenges are missing information... [and finding] the relevant accounts.'' 
[P4] 
even outlines a process wherein they go back to individuals who have responded on social media and actively solicit their locations to address this issue: ``People wouldn't put their location... [and we want to make] sure that we're flagging those, so that we can go back and reach out to those people and ask them where they are''; but this approach is problematic, as ``[people] don't want to give their information to the government.''

Even when practitioners have identified relevant accounts, cross-checking the correctness or timeliness of information from those accounts remains a difficult proposition.
As [P7]  
mentions, determining ``what's useful and what's people exaggerating things'' takes substantial effort, a sentiment also echoed by [P9] 
and [P6] 
across many comments about data validation. 
\inblue{This is in consonance with the fact that lack of human resources and/or management support was cited in our survey as the most important barrier to social media adoption.} 
As [P6] notes, ``for the State agencies in the cities, we have to confirm that the information that we are getting on social media is accurate before we can display it out there''. 
[P4]  
further reports experiences with individuals online responding to questions with incorrect information, further complicating the collection of reliable information.

\paragraph{Collecting Across a Fragmented Space.} These issues are also further complicated by increasing fragmentation in the social media space. 
While we discuss the impacts of fragmentation on dissemination above, this issue spills over into collection, as practitioners know their focus on a small set of platforms biases the data.
As [P5]  
describes, ``people's viewpoint determines which platform they use... [and if an] agency or a jurisdiction has a account on one platform, does it look like we're favoring one party versus the other? You know it's got very complicated recently.'' 
Practitioners also highlight concerns around more-private platforms like WeChat, WhatsApp, and Nextdoor, all of which contain important information but are difficult to collect data from. For instance, [P3] 
mentioned, ``we did have a Nextdoor account, and we would solicit people to comment in Nextdoor. We would get some, but it's hard to search Nextdoor also''. 

\paragraph{The Utility of Multimedia.} Practitioners' descriptions of data collection are not universally negative, however, as a new element in the search for reliable information does emerge in the interviews though, specifically around multimedia. 
Multiple interviewees refer to images and videos 
as being particularly useful. 
For example, [P7]  
 explains how ``videos are useful because you can actually see the water depths [and] you can see if it's going in people's houses.'' 
[P4]  
similarly describes using visual media to assess damage: ``We use a system ... to pull up... pictures and videos, and there were a lot more places that were flooded then were being reported.''

\subsection{Engage-and-Mobilize Practices}
\label{sec:findings-focus-mobilize}


According to the survey results, a gap remains between daily or weekly use of social media 
in 
a personal capacity (85\%) compared to professional use (75\%). Furthermore, the use of social media in EM agencies was sometimes described as ``reactive.'' This may suggest that the professional use of social media by EM practitioners to engage with the public might 
be challenged, perhaps due to the limited human resources. 
Therefore, during the interviews, we sought more information on two-way communication with the public to better understand both the current and the desired mechanisms to facilitate it. 

\paragraph{Interactive Emergency Communication.} While 
practitioners at EM agencies have long interacted with the public via online spaces, much of the practice is around dissemination. The collection in these spaces has been serial and synchronous in nature. 
Practitioners would broadcast information to the general public, then--after enough time to allow the public to respond or share other relevant content--practitioners would sample these spaces, update their internal situational awareness, and determine new information to broadcast ~\cite{hughes_evolving_2012}. 
This process would repeat throughout an event's progression, 
with each pass taking hours or days. 
Shortening this turnaround time for crisis communication has been observed in recent studies pointing towards interactive communication 
~\cite{10.1109/asonam.2018.8508709.2018,zade_situational_2018}, 
such as by automatically identifying requests that practitioners can service. 
Trends in broader studies of crisis communication have similarly moved 
towards interactive communication \cite{10.1111/1468-5973.12130.2018}. 

We see strong evidence of this shift to interactive, parallel communication in our interviews with practitioners as well. 
For example, [P4]  
describes actively building audiences via social media as an avenue for emergency preparedness, saying, ``we [practitioners] also communicate out to [vulnerable audiences] on a regular basis... on how they can prepare,'' or making ``sure that [caregivers] have an emergency plan'' for current emergency. 
Likewise, during wildfires in Turkey, 
[P11]  
describes efforts at active community management, where they received ``20,000 comments from people into asking us to send help, or then once we did, thanking us for sending help.'' 
[P8]  
further describes soliciting reports of impact and missing people from audiences during a cyclone in Mozambique, where they ``used all sorts of social media channels to reach out to people that were on the ground... get feedback... [and] assure people that the majority of people on the missing-people list were found.''

\paragraph{Social Media as a Practitioner Performance Indicator.} 
An implication 
of this more interactive engagement with the public is that the \emph{metrics} around engagement have become increasingly valuable to practitioners. 
Prior studies on 
emergency communication evaluations focus more on practitioner-oriented metrics such as 
\inbrown{the} number 
of partners and message timeliness, with some audience-focused metrics such as 
news coverage or trust assessment \cite{10.2105/ajph.2017.304040.2017}. 
Social media metrics, however, provide a fundamentally more direct method for measuring audience interaction, giving practitioners a new avenue to evaluate their own effectiveness in engaging the public.
Studies as recent as 2018 point to this potential, saying, ``social media can be important at 
in several places in the emergency risk communication process, which may point to the need to develop enhanced social media capacity'' \cite{10.1089/hs.2018.0020.2018}. 

We find confirmation and adoption of this potential, as social media metrics give practitioners insight into impact, reach, audience size, response time, etc. 
For example, [P8]  
points to ``followers'' as a key metric of success for their organization, saying, ``Our followers grew quickly... We had more follow us than national civil protection.'' 
[P4] 
provides a more specific description as well: 

\begin{displayquote}
\emph{[P4]}  
So we measure the number of fans, the number of followers, post engagements. How many posts we do. and then we try and kind of analyze that to see like we can produce a heat map on it, to see like where exactly those fans are coming from.
\end{displayquote}

\paragraph{Social Media for Mobilizing the Public.}  
The new approaches to public engagement by EM agencies also introduce new opportunities for the public to take an active role in the 
response phase of the EM lifecycle. 
\inblue{These new opportunities are apparent in participants' discussions of mobilizing the public to take action via unique affordances of today's social media environment, especially with respect to visual media: e.g., [P2] states, the ``ability to show images and video as opposed to just plain text adds to the level of urgency and the announcement.''}
\inblue{Survey respondents also did not consider that the response phase was their primary focus when engaging with social media, which further increases the need for additional resources for managing social media during this phase.} 
%
Beyond simply reporting, the dual rise of the access to information on social media at large scale and volunteer organizations has led to new opportunities for members of the public to participate as voluntary members of 
community organizations that are involved in various activities across the EM lifecycle 
\cite{peterson2015_iaem,10.1145/2441776.2441832,hughes_social_2015}. Examples include VOST organizations in Europe 
 \cite{vost_europe} 
and 
CERT organizations in the U.S. 
\cite{cert_US}. 

VOSTs are a concept that has been integrated into the EM lifecycle in various capacities. VOSTs emerged in 2011 \cite{st_denis_trial_2012}. They primarily focus on leveraging digital tools and social media to support emergency response efforts, often operating remotely and in a more ad-hoc or flexible manner. Their integration into 
structures of EM processes can vary widely depending on the organization and the specific needs of the community or incident \cite{fathi_vost:_2019}. 
On the other hand, CERTs are a more mature and structured component within the framework of the Citizen Corps, which
 is part of the Federal Emergency Management Agency (FEMA) under the U.S. Department of Homeland Security (DHS). \cite{peterson2015_iaem} CERTs have a standardized training curriculum and are typically more integrated into local EM processes and response plans. Their roles and responsibilities can vary by location, but they generally include a broad range of activities from disaster preparedness education to support in actual emergency situations. 

Considering especially advance-notice disasters, EM agencies could engage the public through either direct calls for action, or more formal activation through these organizations. 
%
Despite early skepticism among practitioners, 
as pointed by 
[P8],  
these digital newcomers have, as shown in several cases, established themselves as valuable resources for EM agencies. 
Online social spaces like Twitter/X and Instagram provide a space for these organizations 
to prove themselves, build credibility, and demonstrate to official 
agencies that they are legitimate and useful actors.

\paragraph{Challenges in Moderating Online Emergency  Communication.}  
While these new interactive engagements with the public come with many advantages, the speed and direct communication introduce new challenges. 
Issues of content modality and platform fragmentation, as discussed above, practitioners also highlight the interactive and public nature of social media-based emergency communication as problematic for content moderation. 
\inblue{This capacity gap might be another driver for the importance that survey respondents gave to ``lack of human resources and/or management support'' as a barrier to social media adoption.}
In particular, as members of the public can directly respond to and comment on practitioners' messages in public forums (e.g., comments 
on Facebook/Instagram/YouTube or replies on 
Twitter/X), policies and capabilities around how to handle unhelpful, incorrect, or misleading information have introduced unique issues. 
As practitioners engage on social media platforms outside 
their own control, these practitioners cannot always delete or hide instances of negative commentary, and some organizational policies may actually forbid such action. 
[P3] 
and [P4] 
make this point clearly. For instance,   

\begin{displayquote}
\emph{[P4]}  
I will say one of the things that doesn't work well is when you give a post that you want people to answer, but people don't answer with... correct information. [People] put completely inappropriate pictures [in response to our messages], and ... we can't stop that. We can't delete it. You know I had someone that recently posted a really graphic YouTube video, and it was just like showing effects of like radiation. I don't know why they posted it on the post, but it was something that I couldn't take down, and it showed up on there as like the one and only comment for weeks... And because of our policies and on Twitter, you can't delete anything... so it just stays there. So the only thing that we can do to kind of counteract that is, keep posting good messages and making sure that kind of goes down on the list, and it's not popping up on people's feeds... There's very strict guidelines on what we can do. 
\end{displayquote}

\subsection{Practitioner-Cooperation Practices}
\label{sec:findings-focus-cooperation}

\inblue{Table \ref{tab:framework_history}'s remaining process concerns cooperation, particularly between EM organizations.} 
\inblue{Though cross-agency organization and cooperation \inbrown{have} 
been studied in prior work~\citep{10.1002/pa.1967,10.1016/j.giq.2020.101539}, much of that work has focused on communication between public agencies, even though FEMA's ESF policies include explicit instruction on engaging with local, community-based and civil-support organizations--e.g., ESF 6 directs EM personnel to engage with local churches.} 
\inblue{In this context,} we identify three major findings related to how EM practitioners and their organizations \inblue{leverage social media to communicate and the barriers they continue to face}.  

\paragraph{Social Media as a Supplemental Cooperation Venue.} 
While prior studies have shown that 
EM agencies learn about new developments from 
partner organizations 
via public social media, a clear consensus from our interviews is that while such cross-practitioner engagement is valuable, \inblue{it remains rare and ad hoc.} 
Numerous interviewees have made it clear that much of the interaction 
between organizations occurs via private channels or backchannels, as [P3]  
and [P8]  
both state: 

\begin{displayquote}
\emph{[P3]}  
``We don't tend to talk to other emergency management agencies via social media. But I know that all of us tend to probably keep an eye on our each other social media.''
\end{displayquote}

\begin{displayquote}
\emph{[P8]}  
``We are in constant communication with our colleagues from other international teams. We're also in permanent contact with our national agencies... [But] we don't use social media to communicate with the official agencies [beyond] private messages. So we do use social media during emergencies, but with official agencies, not publicly'' 
\end{displayquote}

\inblue{Even with back-channel communication, barriers in supporting cross-agency cooperation arise from structural differences across local organizations as well. 
Some counties are more technologically sophisticated than their neighbors, while others have explicit policies in place that forbid engagement via online social spaces, with some jurisdictions relegated to  fax machines for inter-agency reporting. 
These policies make public-private cooperation particularly difficult, as private, citizen-led initiatives are then actively walled off from inter-agency communications, even if they have credibility, resources, and useful information.}

\paragraph{Building Relationships through Social Media Credibility.}  
\inblue{Despite barriers between formal public agencies and among informal public/private EM organizations~\cite{purohit2014identifying}, social media and online social spaces have empowered public, citizen-lead organizations to} 
amass substantial audience--on occasion, larger than even official governmental EM agencies. 
These non-governmental organizations can then establish regular and normalized relations with their local, state, and national EM agencies. 
As [P8] notes:  

\begin{displayquote}
\emph{[P8]}  
``Our first challenges were in the very beginning of our existence... We weren't really known, [and] we had no official government agencies that... had our back... we had to make our way with... our persistence [and] credibility.''
\end{displayquote}

While public agencies might initially be hesitant to engage and cooperate with these private organizations, awareness among the agency's constituency can drive additional political will to engage with these groups.
Barriers in this space are often described as particularly difficult in going from private groups to public agencies. Interviewees from public agencies described partnerships with local citizen organizations (e.g., Black churches in the US during COVID \cite{brewer:dis:2020,ijerph19158926}), while other interviewees housed within private, citizen-led organizations had to either partner with stakeholders already embedded in the public EM agency space or build sufficient audience to open doors to public agencies. 

%
\paragraph{Leveraging Relationships for Social Media Processing.} 
\inblue{This potential for and barriers hindering public/private cooperation are particularly important in today's information ecosystem, as the limited nature and high cost of human resources are well-known challenges} for official EM agencies around the world~\cite{hughes_evolving_2012, hiltz_exploring_2020}.
\inblue{Our survey respondents echoed this sentiment as well, as agencies are consistently under-resourced}.
These constraints limit agencies' abilities to process large-scale social media data for collection practices. 
Community and volunteer response organizations provide a path to mitigate these constraints by offloading some of this collection to trusted volunteers. 
This cooperation with trusted groups facilitates a trustworthy solution for official governmental EM agencies to effectively leverage social media: 
 
\begin{displayquote}
\emph{[P6]}  
``So the CERT members take pictures of what actually is going on, and that information gets fed back to their fire departments. The fire departments disseminate that information to the Department of Transportation [for decision support].'' 
\end{displayquote}

\paragraph{Examples of Public-Private Partnerships.} 

\inblue{
Several examples of public/private cooperation emerged from discussions with interviewees and stakeholders: One such example is how VOST Portugal \cite{vost_portugal_fire} 
coordinated with local and national EM agencies during the 2022 fire season. Their activities included sharing real-time information, 
monitoring fire incidents, and assisting in resource allocation, demonstrating a key role that social media can play in facilitating collaboration and building trust between volunteer organizations and official agencies. 
In addition, during VOST Portugal's involvement in the aftermath of the Turkey-Syria earthquake \cite{vost_portugal_turkey_earthquake}, 
the VOST supported geo-locating requests-for-help using social media, which enabled additional scalable direction of emergency services to affected areas. This rapid response capability demonstrates the value of volunteer organizations in crisis situations. 
In the U.S., the collaboration between the Montgomery County, MD CERT \cite{mc_cert_US} 
and the Qatar Computing Research Institute (QCRI) during Hurricane Dorian~\cite{Imran_Qazi_Ofli_Peterson_Alam_2022} in 2019 also demonstrated the utility of integrating advanced technology with grassroots efforts in disaster response. CERT volunteers used social media platforms to gather and verify real-time data, which QCRI then processed with advanced analytics tools. 
This partnership underscores the importance of trust and verification processes and provides an example of how such collaborations can enhance emergency management practices. 
Finally, the COVID-19 pandemic saw cooperation between academic institutions and emergency management agencies, exemplified by the Montgomery County CERT's collaboration with George Mason University, Brigham Young University, and the University of Texas at Austin~\cite{osti_10391398}. This partnership focused on developing models to track risky behaviors for virus spread using social media data, enabling faster and more accurate public health responses. By actively participating in the project's development and implementation, practitioners helped establish the credibility of social media in disaster response. This collaboration highlights the potential of academic-practitioner partnerships and the role of relationship building and trust through verified and credible social media data in managing crises effectively. 
}

%% file: 06-discussion.tex

\section{Discussion} \label{sec:discussion}

We can draw some insights from the survey and interviews about the challenges and opportunities faced by EM 
practitioners when using social media platforms for communication and engagement with the public.
We begin with a discussion of how social media use in EM 
has changed over time, along with new developments revealed by this research. We then turn our attention to implications for policy and procedure, technology, and socio-technical design.

\subsection{Evolution of Social Media Use}

\begin{figure}[t] 
\centering
\includegraphics[clip, width=\textwidth]{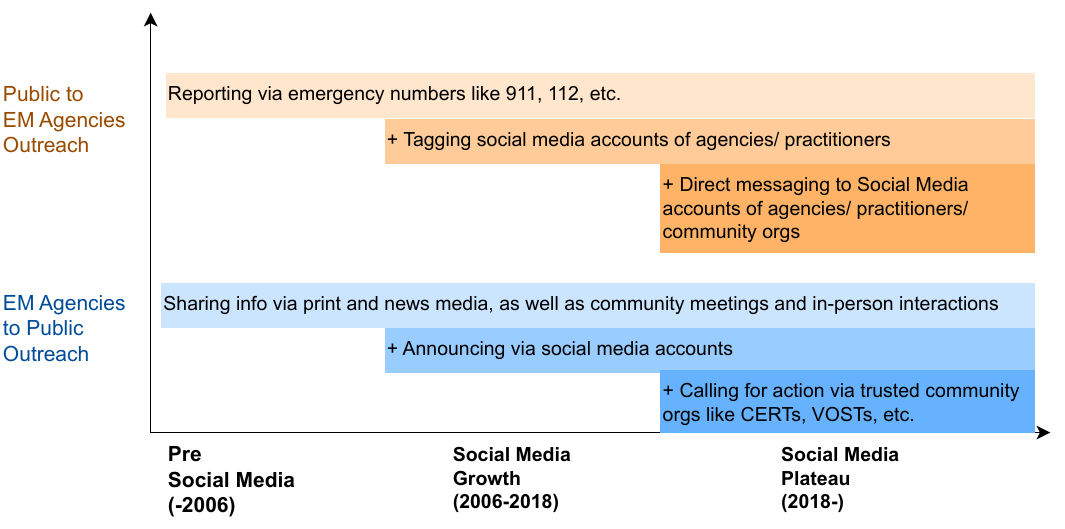}
\caption{Evolution of engagement between EM agencies and the public before the period of popularity and growth of social media platforms and after their plateau of growth, based on social media usage worldwide \cite{social_media_stats}.}
\label{fig:social_evolution}
\end{figure}

The use of social media platforms has increased over the past two decades and now seems to have reached a plateau. For instance, the time that people spend on social media grew from about 90 minutes per day in 2012 to about 140 minutes per day in 2018, but has not experienced substantial change in the last five years~\cite{gwiDigital2023}.
\inblue{EM practitioners have adopted social media almost universally for personal use and, to a large extent, also in a professional capacity, both at the local and non-local levels.} 
%
Engagement practices between the public and EM agencies have taken many different forms before social media, during the social media growth period, and as social media use has plateaued. 
As shown in Figure~\ref{fig:social_evolution}, EM officials communicated with 
the public primarily using print and traditional broadcast news media, as well as in-person community meetings and interactions before the introduction of social media. 
However, this approach often resulted in delayed messaging at a time when the public needed timely updates on 
an ongoing disaster. 
Social media offers a way to rapidly broadcast relevant information and engage 
the public. Hence, as such platforms gained popularity, EM agencies gradually increased their social media presence for public communication. 
In interviews, several participants remarked that social media is no longer considered a ``new''  medium, but rather an important part of their communication toolbox. This observation marks a significant shift in attitudes toward social media.
Social media use has further progressed to involve collaboration between 
EM agencies and trusted community volunteer organizations, such as VOSTs and CERTs, to manage social media communications during disaster events. 
These digital volunteer organizations help amplify and ensure the dissemination of messages to the public, including situational updates and calls for action.  
While past research ~\cite{hughes_social_2015, st_denis_trial_2012, fathi_vost:_2019} has discussed cooperation with these volunteer organizations, our findings suggest that EM agencies have only recently started to collaborate more with digital volunteer organizations.

Similarly, the ways in which the public can communicate with official EM agencies have changed with the widespread use of social media.
Before the era of social media, the public would call emergency numbers (e.g., 911 in the U.S. or 112 in Europe) or speak directly to officials to report emergencies or seek assistance. 
These practices have since expanded to include tagging social media accounts of EM agencies for reporting events and seeking help~\cite{10.1109/asonam.2018.8508709.2018}, especially as more EM agencies establish official accounts on popular social media platforms such as Twitter/X and Facebook. 
Further, as the social media policies of the EM agencies evolved to include elements of ``social listening,'' the public has increasingly reached 
out to EM agency accounts and known officials through direct messaging. 
Our research finds that EM agencies now engage in more interactive and parallel communication with the public through social media~\cite{10.1109/asonam.2018.8508709.2018,vitiugin2024multilingual}, marking a significant departure from their previous usage of these platforms. 

An emerging challenge highlighted by this research is the fragmentation of public attention across a diverse range of social media platforms. 
In the initial surge of social media popularity, a significant concentration of users could be found on the most widely used platforms, allowing EM agencies to reasonably engage with a large portion of their constituents by focusing on a few key platforms. 
This is no longer the case. 
Currently, public attention is dispersed across many platforms, influenced by factors such as age, demographics, geographic location, interests, and cultural 
norms. 
This shift presents a challenge to EM agencies as they seek to communicate with the public in a way that is inclusive and reaches all the individuals affected by a disaster event. 
It also complicates the task of \inblue{(mostly incident/event-specific)} monitoring and responding to social media activity spread across multiple platforms. 
Our data reveals that EM agencies have started relying more on volunteer organizations, as well as individuals or influencers with a broad social media audience,  to address this problem. 
This reliance also opens new opportunities at the intersection of EM and public health, where community-based organizations have emerged as trusted messengers in efforts to increase vaccination rates \cite{SHEN20231994}.
Clearly, the complexity of this challenge is expected to persist and continue to grow over time. 

EM agencies have faced difficulties in monitoring social media since the advent of its use by the public, including challenges related to \inblue{capacity and lack of resources, particularly in relation to} the volume and veracity of information~\cite{hughes_evolving_2012,tapia_seeking_2011}. 
These challenges have only exacerbated over time~\cite{hiltz_exploring_2020,lorini2021social}. 
The fragmentation of public attention (discussed above) has led to an increase in the number of platforms to monitor and a surge in the volume of information. Concerns about 
the generation of inaccurate, untimely, or deliberately misleading information on social media have escalated in recent years ~\cite{allcott_trends_2019,muhammed_t_disaster_2022}. 
The COVID-19 pandemic, in particular, witnessed a significant rise in concerns about the trustworthiness of information disseminated through social media ~\cite{ferrara_misinformation_2020, cinelli_covid-19_2020}. Our data aligns with these concerns. While much research has been conducted to address information overload for emergency managers during disaster events ~\cite{palen_social_2018, 10.1109/asonam.2018.8508709.2018, 10.1145/2771588.2015} and to identify false and misleading information in social media \cite{hunt_monitoring_2022, arif_how_2016, mehta_trust_2017}, these problems are unsolved and we anticipate they will remain significant concerns in the years ahead.

Finally, our survey findings show a higher number of practitioners using social media during the mitigation and preparedness phases of the EM lifecycle, followed by the response phase. This suggests that the acceptance of 
social media extends beyond the response phase. While extensive research literature has focused on response \cite{reuter.iscram:2018}, including development of tools and analyses in the aftermath of disasters, these findings demonstrate a growing interest and need to research effective technologies and tool designs to support the social media use in mitigation and preparedness phases.  

\subsection{Policy and Procedure Design 
Needs} 

In analyzing the diverse challenges interviewees reported on social media use within their agency, we have identified 
a common theme 
-- lack of procedures and policy implementation for both social listening and community engagement. 
We present the following policy and procedure needs 
for practitioners.  

\paragraph{Updating Communication Plans} 
These results demonstrate that EM agencies need to develop a policy 
to formally incorporate social media platforms within their communication plans for 
engaging with their community members 
\emph{during all phases of the EM lifecycle}. 
While this need may be obvious, we note that at least one interviewee, [P6], expressed frustration that their state-level policies disallowed the use of social media; and other respondents described a more ad-hoc approach to these spaces. 
Such policies would meet the expectation indicated by over 90\% 
of the survey participants 
that social media could provide somewhat high or very high value for public communications. 

\paragraph{Supporting Engagement across the EM Lifecycle} 
Standard procedures are needed for training practitioners responsible for public communications (e.g., Public Information Officers) 
in using social media 
throughout all phases of the EM lifecycle. 
\inblue{Given that many EM practitioners take multiple roles in their organization, as per the survey results, this training would probably need to be provided to people in roles that might not be directly related to public communications.} 
[P4]'s interview responses demonstrate the utility of this enhanced engagement for building larger audiences during blue-sky days prior to emergencies, understanding reach, and having a better understanding of who their audience is--alleviating some concerns around trust and geolocating accounts. 
By looking at public communications holistically across all phases, EM agency readiness, from a communication perspective, would be better positioned to effectively engage and mobilize the public via social media, 
before, during, or after disasters. 

\paragraph{Scaling Up via Formal Partnerships with Community- and Volunteer-Response Organizations} 
A critical barrier (observed in the survey responses of Q10.4) to using social media in emergency operations centers is  
the lack of resources, both human and logistic, as highlighted by prior research~\cite{hiltz_exploring_2020}. 
EM agencies can consider developing a Memorandum of Agreement or Memorandum of Understanding with trusted community-/ volunteer-response organizations, such as CERTs or VOSTs, to outsource some parts of the information processes, as described in Section \ref{sec:findings-focus-mobilize} 
above.  

\paragraph{Capitalizing on Existing Transnational Resources} 
Given the guidance needed by local EM agencies to address policy and procedural 
barriers in social media adoption, 
existing resources (e.g., laws, guidance, plans) can be consulted. 
For example, 
the U.S. Department of Homeland Security's (DHS) National Emergency Communications Plan (NECP \cite{NECP_comm_plan})  
could provide guidance on developing emergency communications plans and procedures that address evolving 
technologies to support novel work capabilities, including 
developing and delivering training programs. 
Similarly, the Data Act \cite{data_act_EU} 
adopted by the European Union can be a valuable resource, which aims to enhance the value of data generated by connected products and services. 
For instance, it can allow 
EM agencies 
to access data held by enterprises during natural or man-made disasters, promoting a more informed and efficient response.  The Act also ensures fairness in data-sharing contracts and enhances the accessibility of crucial data through provisions for cloud service switching. 
Additionally, the Digital Services Act \cite{digital_service_act},   
to be implemented in February 2024 by the European Union, is another significant legislation that targets regulating the accuracy of information and curbing the spread 
of misinformation on social media platforms. For EM agencies, this means a more reliable flow of information during disasters, 
which is essential for effective decision making and public communication. 

Through guidance of the NECP, the Data Act, and the Digital Services Act, EM agencies could capitalize on the value social media would bring to their work practices. 
Furthermore, agencies outside of the U.S. and Europe could seek resources developed by the International Telecommunication Union (ITU), which encourages the development of emergency telecommunication plans as part of a country's risk reduction strategy. ITU developed guidelines \cite{itu_guidelines} 
to assist policymakers in using communication technologies and services in all phases of the EM lifecycle. 


\subsection{Technology Design Needs} %

\inblue{Across our survey and interview responses}, we \inblue{have identified several needs for which technological solutions could facilitate and accelerate EM communication processes}. 
\inblue{While our interviews highlight well-known and persistent needs (e.g., geolocating social media content to ensure accurate situational awareness~\cite{maceachren_senseplace2_2011} or calls for multi-platform collection~\cite{10.1126/science.aag2579.2016}), below, we describe new needs that have emerged in response to the modern, complex, and fragmented information environment}. 
\inblue{We emphasize that these technological solutions should amplify human capacity rather than replace it; in the context of life-threatening EM, technological solutions must support human-in-the-loop processes to mitigate associated techno-centric risks--e.g., AI hallucinations, biases, etc.--as these risks are increasingly studied in related safety-critical fields~\citep{HOSTETTER2024,OVIEDOTRESPALACIOS2023106244,10.1001/jama.2023.9630,10.1145/3571884.3603756}}.

\paragraph{Managing Multiple \inblue{Audiences across} Social Media Platforms} \inblue{In the halcyon days of social media, the vast majority of the internet audience could be reached across three core platforms--YouTube, Facebook, and Twitter/X.} 
\inblue{In the late 2010s, however, many new spaces began to emerge~\cite{2024:pew:gottfried}, which has \inbrown{led} 
to a fragmentation of social media audiences.}
\inblue{This fragmentation substantially hinders reaching heterogeneous audiences for EM practitioners, as younger audiences now prefer Instagram, older audiences prefer Facebook, while Hispanic and Black \inbrown{people} are over-represented on TikTok~\cite{2024:pew:gottfried}.} 
\inblue{The proliferation of hyper-local platforms like Nextdoor~\cite{Brown_Sanderson_Graham_Kim_Tucker_Messing_2024} and politically partisan/alternative platforms like Gab~\cite{Fair_Wesslen_2019,10.1177/14614448211045662} further exacerbate this issue for EM practitioners}.
\inblue{While pushes to \emph{collect} information across platforms are not new in the EM space~\cite{kaufhold2020mitigating,hughes_information_2016}, the need to manage multiple audiences \emph{across} these platforms is a problem unique to the current fragmented space, where politically liberal users are unlikely to be on right-leaning partisan platforms, and older individuals are less likely to be on Instagram.}
\inblue{This audience fragmentation} requires analytics capabilities for EM agencies to better understand the various types of audience \inblue{across these} platforms. 
\inblue{This audience understanding represents a new priority we see among practitioners as social media spaces have become normalized communication channels:} a focus on \emph{inclusivity} in disseminating information across platforms to reach all population groups, \inblue{especially hard-to-reach audiences}.  
\inblue{This prioritization has become especially clear in the aftermath of the COVID-19 pandemic, where minoritized groups, which comprise the majority of essential workers, were particularly disadvantaged~\cite{10.1145/3555197}}. 
\inblue{These needs} represent an area with potential for new technology designs in supporting multi-platform monitoring and engagement.

\paragraph{Supporting Multimodal \inblue{{Collection and Dissemination}}} 
Practitioners have frequently identified the need for multimodal (i.e., combining two or more data modalities of text, audio, image, and video) messaging support in public communication for inclusiveness, as well as multimodal data collection for enhanced situational awareness~\cite{lorini2021social}. 
The processing, integration, analysis, and recommendation of relevant information from multimodal data on social media platforms require advanced technological capabilities. 
Collecting and synthesizing multimodal data remains technically challenging but has received much research attention recently~\cite{imran_using_2020}.  

\inblue{EM practitioner needs in this space are unique in that these needs cover \emph{both} needs to support the dissemination of visual media with text \emph{and} the analysis of multimodal data.} 
\inblue{As we discuss in \S5.2, multiple practitioners 
find the value in using visual media to enhance situational awareness, but also in \S5.3, practitioners mention how much more effective visuals are for informing and mobilizing the public}. 
\inblue{Collecting, storing, and analyzing visual media is a computationally intensive process that can take orders of more magnitude resources than storing text~\cite{weber2020eccv}, and few tools exist to support EM practitioners' needs to \emph{search} this visual media for response, e.g., assessing damage.} 
\inblue{Simultaneously, EM practitioners understand the utility of visual media in communicating to general audiences, and tools to support rapid creation of infographics--which are particularly popular during disasters~\cite{2022:buntain:iscram}--that are tailored for specific platforms would accelerate this dissemination.} 

\inblue{Recent advances in generative AI models, including for visual media like DALL-E, multimodal models like GPT-4o and language-vision models like BLIP~\cite{pmlr-v202-li23q} may be especially valuable in this dual-use collection/ dissemination role for EM practitioners}. 
\inblue{At the same time, substantial care must be taken in general application of these generative AI models, as they are not without their risks~\cite{10.1145/3571884.3603756,10.1001/jama.2023.9630}.} 

\paragraph{Facilitating Multiple Languages}  The increasing number of disasters around the world in geographical areas with distinct social, cultural, and linguistic characteristics demand effective technologies in multilingual settings. 
Recent advances in tools to support EM agencies 
through language technologies have primarily been demonstrated for English text ~\cite{reuter.iscram:2018,palen_social_2018}.
EM agencies in various parts of the world, including large and densely populated urban regions with diverse populations, would like to reach these diverse audiences effectively in times of a disaster. 
Thus, there is a need for cross-lingual, cross-region, and cross-hazard analytical models to support EM agencies.  
Prior work has shown success in leveraging cross-hazard data \cite{2021,vitiugin2024multilingual}, and recent machine translation models can similarly support content creation and dissemination, 
as well as understanding across language barriers. 

\paragraph{Summarizing Online Data across Platforms} The volume of data produced across multiple social media platforms requires the ability to summarize content across these platforms. 
The challenges of volume, velocity, and veracity for big crisis data are well recognized in the crisis informatics literature \cite{castillo2016big,lorini2021social} that motivate this need. The technological capabilities 
for how to collect and align information across platforms and synthesize effective summaries \cite{mccreadie2023crisisfacts} 
are relatively new for social media analytics for EM agencies. 

\paragraph{Opportunities and Hazards around AI}
We note that many of our interviews occurred prior to or in the early days of massive news coverage around AI advances in Large Language Models (LLMs) and the widespread availability of tools like ChatGPT \cite{chatgpt}. 
%
LLMs and related generative AI models can support content creation, content summarization, multilingual processing, multimodal processing, conversion across modalities (including video-to-text, image-to-text), and so on. 
While rigorous scientific investigation into the impact these AI tools will have on EM practices is currently in its nascent stages, early work makes it clear that these tools are already finding use within the EM world, as \citet{HOSTETTER2024} demonstrates with ``this technology will likely be elemental to our engineers'  practice and education.''
While this integration may 
show 
many open needs around cross-language translation 
and 
multimodal generation/analysis, substantial 
barriers 
exist around information quality, bias, and a variety of safety concerns that warrant attention~\cite{OVIEDOTRESPALACIOS2023106244}. 
Also, they 
demand more co-designing efforts and engagement between technical researchers and EM practitioners.~\cite{JRC_report}



\subsection{Socio-Technical Design Needs} 

Though practical and political barriers have complicated practitioners' use of social media for 
emergency communications and operations of EM agencies, 
our work has found new uses and barriers that have arisen at the intersection of the social and technological aspects surrounding social media. 
These socio-technical needs fall into two main areas: 1) better empowering practitioners to engage with their audiences, and 2) better supporting multi-party, transnational engagement among practitioners, volunteers, and other stakeholders. 
Unlike many historical solutions to problems in crisis informatics, solutions to these needs transcend technology, cannot be solved with policy alone, and require a degree of social understanding and integration.

\paragraph{Empowering Practitioners to Moderate their Spaces} A clear example of a need at this intersection concerns practitioners' abilities to moderate their social media communication channels.
Multiple interviewees recounted stories of incorrect, low-quality, or otherwise disturbing content that individuals posted to EM agencies' 
accounts/posts/channels. 
Common among these stories is that practitioners are ill-equipped to respond to this low-quality content, either because local government policy considerations preclude deleting said content, or the social media platforms themselves do not empower targeted moderation. 
As an example, if a bad actor posts a message on Twitter/X in reply to a practitioner's post, the practitioner cannot delete it, even if that message contains incorrect or harmful information. 
While larger adoption of community-moderation capabilities like those on Reddit could better empower practitioners to delete misinforming content, lock threads, etc., policy may still preclude such responses. 
Hence, practitioners are in need of new spaces that simultaneously enable engagement with the broader social audience while empowering practitioners with control over the content their audiences see. 

\paragraph{Inclusive Messaging in a Fragmented Ecosystem} While the fragmentation of this broader audience is itself an issue, practitioners note the difficulty that comes with engaging across the many social media platforms and modalities available; e.g., Facebook, Twitter/X, Nextdoor, YouTube, TikTok, Instagram, etc. 
Technological solutions that expand covered platforms, while useful, do not solve this fragmentation problem, as key open questions exist around 
which segments of the audience 
use these platforms. 
Practitioners understand that different platforms provide access to different age groups, different language communities, and other segments of their audience, but practitioners have few tools that help them understand the social gaps in the data they collect or audiences they desire to reach. 

\paragraph{Integrating Diversity in Metrics of Engagement} This question of reach is of particular value to practitioners as we show they are increasingly relying on social media's engagement metrics to assess how well they are communicating with the public. 
As such, socio-technical solutions that both reveal which segments of the audience are reached and which segments are engaging with practitioners would better support practitioners in ensuring a more equitable and effective public information campaign, as practitioners could quickly assess which audiences are responding versus which are silent. 

\paragraph{Mobilization and Self-Determination} This point of silent audience segments is important to note here as well because one of the major capabilities that social media provides is more direct engagement with and mobilization among the general public. 
This capability could translate into enhanced self-determination and self-advocacy among audiences engaging with practitioners. 
Tools that help practitioners identify and engage with influential and trusted members among these audiences and bring these individuals into the decision-making and communication process could vastly improve and accelerate dissemination and collection of critical, high-priority information. 

\paragraph{Supporting Transnational Relief Efforts} The opportunity for self-determination and a more peer-to-peer engagement among practitioners and trusted members of the public can extend beyond regional borders.
As [P8] 
describes, volunteer organizations--i.e., not explicit practitioners--can provide utility to less well-resourced communities, but social barriers around credibility must first be overcome or local connections forged. 
Online social platforms can bring the resources of these remote but well-resourced organizations to 
locally impacted groups. 
Particularly enterprising groups have already found success in these transnational spaces, as seen in the comments by [P8]  
about their organization's activation during a cyclone in Mozambique and the 2010 Haiti earthquake \cite{10.1145/1978942.1979102}, but these groups succeed in spite of socio-technical barriers of the modern information environment.